# Medical deep learning—A systematic meta-review

Jan Egger[a,b,c,d,e,*], Christina Gsaxner[a,b,c], Antonio Pepe[a,c], Kelsey L. Pomykala[d], Frederic Jonske[c,d], Manuel Kurz[a,c], Jianning Li[a,c,d], Jens Kleesiek[d,e,f]

[a] Institute of Computer Graphics and Vision, Faculty of Computer Science and Biomedical Engineering, Graz University of Technology, Inffeldgasse 16, 8010 Graz, Styria, Austria
[b] Department of Oral &Maxillofacial Surgery, Medical University of Graz, Auenbruggerplatz 5/1, 8036 Graz, Styria, Austria
[c] Computer Algorithms for Medicine Laboratory, Graz, Styria, Austria
[d] Institute for AI in Medicine (IKIM), University Medicine Essen, Girardetstraße 2, 45131 Essen, Germany
[e] Cancer Research Center Cologne Essen (CCCE), University Medicine Essen, Hufelandstraße 55, 45147 Essen, Germany
[f] German Cancer Consortium (DKTK), Partner Site Essen, Hufelandstraße 55, 45147 Essen, Germany

ABSTRACT

Deep learning has remarkably impacted several different scientific disciplines over the last few years. For example, in image processing and analysis, deep learning algorithms were able to outperform other cutting-edge methods. Additionally, deep learning has delivered state-of-the-art results in tasks like autonomous driving, outclassing previous attempts. There are even instances where deep learning outperformed humans, for example with object recognition and gaming. Deep learning is also showing vast potential in the medical domain. With the collection of large quantities of patient records and data, and a trend towards personalized treatments, there is a great need for automated and reliable processing and analysis of health information. Patient data is not only collected in clinical centers, like hospitals and private practices, but also by mobile healthcare apps or online websites. The abundance of collected patient data and the recent growth in the deep learning field has resulted in a large increase in research efforts. In Q2/2020, the search engine PubMed returned already over 11,000 results for the search term 'deep learning', and around 90% of these publications are from the last three years. However, even though PubMed represents the largest search engine in the medical field, it does not cover all medical-related publications. Hence, a complete overview of the field of 'medical deep learning' is almost impossible to obtain and acquiring a full overview of medical sub-fields is becoming increasingly more difficult. Nevertheless, several review and survey articles about medical deep learning have been published within the last few years. They focus, in general, on specific medical scenarios, like the analysis of medical images containing specific pathologies. With these surveys as a foundation, the aim of this article is to provide the first high-level, systematic meta-review of medical deep learning surveys.

## 1. Introduction

Deep learning [1] had a remarkable impact on different scientific fields during the last years. This was demonstrated in numerous tasks, where deep learning approaches were able to outperform the standard methods, including image processing and analysis [2,3]. Moreover, deep learning delivers reasonable results in tasks that could not have been performed automatically before, like autonomous driving [4,5]. There are even applications where deep learning outperformed humans, like in object recognition [6] or games [7,8].

A field in which this development has begun to show huge potential is the medical domain. With the collection of large quantities of patient records and data, and a trend towards personalized treatments, there is a great need for automatic and reliable processing and analysis of this information [9]. Patient data is not only collected in clinical centers, like hospitals and private practices, but also by mobile healthcare apps or online websites. Together this resulted in new, massive research efforts during the last years. In Q2 of 2020, the search engine PubMed returns already over 11.000 results for the search term "*deep learning*", and around

* Corresponding author.
*E-mail address:* egger@tugraz.at (J. Egger).

90% of these publications are from the last three years. However, even though PubMed represents the largest search engine in the medical domain, it does not cover all medical-related publications. For example, medical topics are also covered in primary venues for computer science research, most often conferences [10,11]. Despite their high impact and consideration within the community, conference proceedings are usually not listed under PubMed, with only a few exceptions, like the prestigious annual conference '*Medical Image Computing and Computer Assisted Intervention*' (MICCAI). In addition, there are rather technical, non-interdisciplinary conferences, for example in computer vision, through which very influential research on medical applications is published [12]. These contributions are often overlooked by medical search engines. This does not relate to survey and review articles, which, due to their length, are generally published in peer-reviewed, PubMed-indexed journals. However, for this reason, it is possible for a review article to miss some relevant contributions.

Taking all these considerations into account, a complete overview of the field of medical deep learning is almost impossible to obtain and acquiring a full overview of medical sub-fields becomes increasingly more difficult. Nevertheless, several review and survey articles about medical deep learning have been published within the last years. They focus, in general, on specific medical scenarios, such as the analysis of certain medical images containing specific pathologies, like the automatic detection of a cardiovascular disorder in computed tomography angiography acquisitions [13]. In this context, the aim of this contribution is to provide an introductory, high-level and systematic meta-review of medical deep learning surveys. Modeled after existing meta-reviews in the medical domain, such as the systematic review of systematic reviews of homeopathy [14], or the survey of surveys on the use of visualization for interpreting machine learning models [15] in a technical domain. The authors are not aware of any meta-review in medical deep learning or general deep learning so far. Compared to medicine, which has a millennia-old tradition, computer science is a very young discipline. Nonetheless, if this discipline continues growing at the current pace, meta-reviews like this will become more and more common.

In this publication, we present all review and survey articles published from 2017 to 2019 found by a systematic PubMed search (see Search Strategy). We did not include articles published after 2019, since we also present the citations of the reviewed publications. Thus, the relatively new reviews from 2020 are still 'undercited' in comparison to the reviews from the previous years (as can also be seen in the decreasing number of citations for 2017: 6089 citations, 2018: 947 citations and 2019: 408 citations), and one aim of this contribution is to give an overall impression of the impact these works have already had on their respective scientific fields. Table 1 gives an overview of the number of reviews published each year, from 2017 to 2019. Furthermore, the table shows the sum of the overall references and citations for each year according to Google Scholar (status as of August 2020). Tables 2–4 describe the publications of each year in more detail.

***Systematic literature review phase.*** For our systematic review, we started with planning the overall structure and main headings of this manuscript, orienting on existing surveys and meta-surveys in the literature. Next, we decided on the databases and years of publication that we wanted to include in our meta-survey. While keeping in mind the overall number of publications we want to cover within our manuscript. Subsequently, we performed the final literature search (see next paragraph *Search Strategy*), summarized every survey and extracted the citations, main architectures, evaluations, pros/cons, challenges and future directions. Finally, we analyzed the commonalities and drew a conclusion resulting in a discussion and future outlook.

***Search Strategy.*** For this systematic meta-review, a search in PubMed for the keyword 'Deep Learning' together with any keyword including {'Review', 'Survey'} was performed. Based on the titles and abstracts, all records which were not actually review or survey contributions in the medical field, like [16,17] and [18], or were not written in English, like [19,20], or are veterinarian reviews [21], or are about a human learning strategy called *Deep Learning* [22], were excluded (while the term "*deep learning*" was coined by Geoffrey Hinton in terms of learning deep neural networks in 2006 [23,24], the term seemed to have existed much earlier in educational psychology [25]. Note further, that non-shallow neural networks had already become an explicit research subject by the early 1990s, when they also became practically feasible to some extent through the help of unsupervised learning [26]). This ultimately resulted in a total number of 43 review or survey publications about deep learning in the medical field, which are covered within this systematic meta-review. Summarized, this high-level systematic meta-review provides an overview of the published medical deep learning reviews and surveys in PubMed, as well as their references and citations (status as of August 2020). Note that our systematic search strategy does not cover all topics in medical deep learning, like a survey about uncertainty quantification in deep learning applications in medical data analysis [27]. However, we did not want to "break" our systematic meta-review search by adding what is arguably arbitrary additional literature.

***Manuscript Outline.*** The main body of this contribution presents exclusively reviews and surveys on medical deep learning from a systematic PubMed search. To keep the manuscript concise for the reader, we provide only high-level summaries and excerpts, mainly form the review and survey abstracts (note that some of the presented reviews cover up to several hundred publications themselves). Thus, every review or survey publication will be summarized in around 100 to 200 words. However, by pointing to the associated publications via the keyword classifications and chronological arrangement of the presented medical deep learning reviews or surveys, the interested reader should be able to dive deeper into the specific categories and sub-categories. The rest of this manuscript is organized as follows: Section 2 presents the overview of the medical deep learning reviews and surveys divided into the years of publications from 2017 to 2019 in chronological order, beginning with the first published work in the respective year. The final Section 3 concludes this contribution with a discussion and outlines areas of future directions.

***Research questions.*** The overall aim of this systematic meta-review is to analyze reviews and surveys published between 2017 and 2019 in medical deep learning. In doing so, we defined the following main research questions for our study:

**Table 1**
Overview of published reviews of deep learning in the medical field from beginning 2017 to end of 2019 according to PubMed and number of citations according to Google Scholar (status as of August 2020).

| Year ▼ | Number of publications | Number of references | Citations (until August 2020) |
|---|---|---|---|
| 2017 | 7 | 1060 | 6089 |
| 2018 | 15 | 1684 | 947 |
| 2019 | 21 | 2279 | 408 |
| **Sum** | **43** | **5023** | **7444** |

**Table 2**
List of published reviews of deep learning in the medical field in 2017 according to PubMed and number of citations according to Google Scholar (status as of August 2020); ordered by epub (electronic publication) date.

| Medical field/subject | Publications | Date (epub) ▼ | Number of references | Citations (until August 2020) |
|---|---|---|---|---|
| Medical image analysis (I.) | Shen et al. [28] | March 09, 2017 | 117 | 1232 |
| Healthcare | Miotto et al. [29] | May 06, 2017 | 119 | 624 |
| Medical image analysis (II.) | Litjens et al. [30] | Jul. 26, 2017 | 439 | 3696 |
| Stroke management | Feng et al. [31] | Sep. 27, 2017 | 55 | 40 |
| Analysis of molecular images in cancer | Xue et al. [32] | Oct. 15, 2017 | 60 | 23 |
| Health-record analysis | Shickel et al. [33] | Oct. 26, 2017 | 63 | 377 |
| Microscopy image analysis | Xing et al. [34] | Nov. 22, 2017 | 207 | 97 |
| **Sum** | – | | **1060** | **6089** |

**Table 3**
List of published reviews of deep learning in the medical field in 2018 according to PubMed and number of citations according to Google Scholar (status as of August 2020); ordered by epub (electronic publication) date.

| Medical field/subject | Publications | Date (epub) ▼ | Number of references | Citations (until August 2020) |
|---|---|---|---|---|
| Toxicity of chemicals | Tang et al. [35] | Mar. 01, 2018 | 103 | 13 |
| Pulmonary nodule diagnosis | Yang et al. [36] | Apr. 2018 | 42 | 14 |
| Physiological signals | Faust et al. [37] | Apr. 11, 2018 | 166 | 301 |
| DNA sequencing | Celesti et al. [38] | Apr. 12, 2018 | 52 | 14 |
| Radiotherapy | Meyer et al. [39] | May 17, 2018 | 234 | 86 |
| Ophthalmology | Grewal et al. [40] | May 30, 2018 | 33 | 29 |
| Electronic health records | Xiao et al. [41] | Jun. 08, 2018 | 123 | 146 |
| Bioinformatics | Lan et al. [42] | Jun. 28, 2018 | 127 | 85 |
| Personalized medicine | Zhang et al. [43] | Aug. 07, 2018 | 142 | 8 |
| 1-D biosignals | Ganapathy et al. [44] | Aug. 29, 2018 | 117 | 19 |
| Omics | Zhang et al. [45] | Sep. 26, 2018 | 143 | 40 |
| Sport-specific movement recognition | Cust et al. [46] | Oct. 11, 2018 | 98 | 35 |
| Diabetic retinopathy | Nielsen et al. [47] | Nov. 03, 2018 | 42 | 12 |
| Image cytometry | Gupta et al. [48] | Dec. 19, 2018 | 137 | 46 |
| Radiology | Mazurowski et al. [49] | Dec. 21, 2018 | 125 | 99 |
| **Sum** | – | | **1684** | **947** |

**Table 4**
List of published reviews of deep learning in the medical field in 2019 according to PubMed and number of citations according to Google Scholar (status as of August 2020); ordered by epub (electronic publication) date.

| Medical field/subject | Publications | Date (epub) ▼ | Number of references | Citations (until August 2020) |
|---|---|---|---|---|
| Medical imaging | Biswas et al. [50] | Jan. 01, 2019 | 94 | 28 |
| Brain cancer classification | Tandel et al. [51] | Jan. 18, 2019 | 123 | 33 |
| Electroencephalogram | Craik et al. [52] | Feb. 26, 2019 | 123 | 91 |
| Pulmonary nodule detection | Pehrson et al. [53] | Mar. 07, 2019 | 48 | 21 |
| Neuro-oncology | Shaver et al. [54] | Jun. 14, 2019 | 81 | 9 |
| Diabetic retinopathy | Asiri et al. [55] | Aug. 07, 2019 | 138 | 21 |
| Cardiac arrhythmia | Parvaneh et al. [56] | Aug. 08, 2019 | 20 | 4 |
| Protein structure | Wardah et al. [57] | Aug. 12, 2019 | 72 | 7 |
| Electroencephalography | Roy et al. [58] | Aug. 14, 2019 | 249 | 101 |
| Neurology | Valliani et al. [59] | Aug. 21, 2019 | 83 | 8 |
| Cancer diagnosis | Munir et al. [60] | Aug. 23, 2019 | 167 | 20 |
| Ultrasound | Akkus et al. [61] | Sep. 03, 2019 | 78 | 7 |
| Radiation oncology | Boldrini et al. [62] | Oct. 01, 2019 | 64 | 10 |
| Drug–drug interaction | Zhang et al. [63] | Nov. 04, 2019 | 180 | 4 |
| Urology | Suarez-Ibarrola et al. [64] | Nov. 05, 2019 | 56 | 10 |
| Sleep apnea | Mostafa et al. [65] | Nov. 12, 2019 | 93 | 5 |
| Ophthalmic diagnosis | Sengupta et al. [66] | Nov. 22, 2019 | 123 | 13 |
| Alzheimer's disease | Ebrahimighahnavieh et al. [67] | Nov. 27, 2019 | 201 | 4 |
| Pulmonary nodule detection | Li et al. [68] | Nov. 29, 2019 | 60 | 3 |
| Liver masses | Azer [69] | Dec. 15, 2019 | 45 | 5 |
| Pulmonary medical imaging | Ma et al. [70] | Dec. 16, 2019 | 181 | 4 |
| **Sum** | – | | **2279** | **408** |

1) What are the different applications of deep learning in medicine?
2) What are the methods most frequently or successfully employed by deep learning in medicine?
3) What are the strengths and limitations of these methods, especially with respect to the field they are applied to?
4) What are the key research gaps that are being investigated or should be investigated according to researchers?

## 2. Medical deep learning: a compact overview of reviews and surveys

This section presents an overview of review and survey publications in medical deep learning. The publications are arranged in three sub-sections by their year of publication, from 2017 to 2019. Within the yearly sub-sections, the publications are arranged chronologically by their date of publication starting with the first published work in the corresponding year. Typically, review or survey contributions order the reviewed publications in categories, like medical image classification, object detection, segmentation, registration, and other tasks. However, for this meta-review we decided explicitly for an order by publication date to show the historical sequence in which they occurred to the reader. Still, the tables provide also a quick overview of the different categories. Hence, at the beginning of every section (2017, 2018 and 2019), the areas of the reviews are summarized in a listing, which corresponds to the chronical order of the publications of this year in the following descriptions. Consequently, Tables 1–4 are also divided into the years 2017 to 2019 and chronically ordered. Note, that the reviewed surveys can focus on a specific subject, like the survey about diabetic retinopathy screening, or span over a general field, like the survey about healthcare. Moreover, the tables present the number of referenced works and the current citations for every year and publication according to Google Scholar, which reveals an overall number of 5023 referenced works in the proposed reviews, and an overall number of 7444 citations for the reviews themselves (status as of August 2020). Furthermore, Fig. 1 shows a collage, where we mapped all figures of the reviewed articles to the surface of the left hemisphere of the brain. Finally, and equivalent to [37], Fig. 2 shows a network visualization for the review articles supplied keywords from 2017 to 2019. More specifically, the figure shows the co-occurrence network and the topic clusters for the article keywords, and it reveals the two main clusters, namely “humans” and “deep learning”, and their connections. Further main clusters center around the keywords “machine learning”, “neural networks (computer)” and “algorithms”. Overall, the clusters and connections show how the medical domain has been affected by deep learning in these years, covering a broad range of topics and applications.

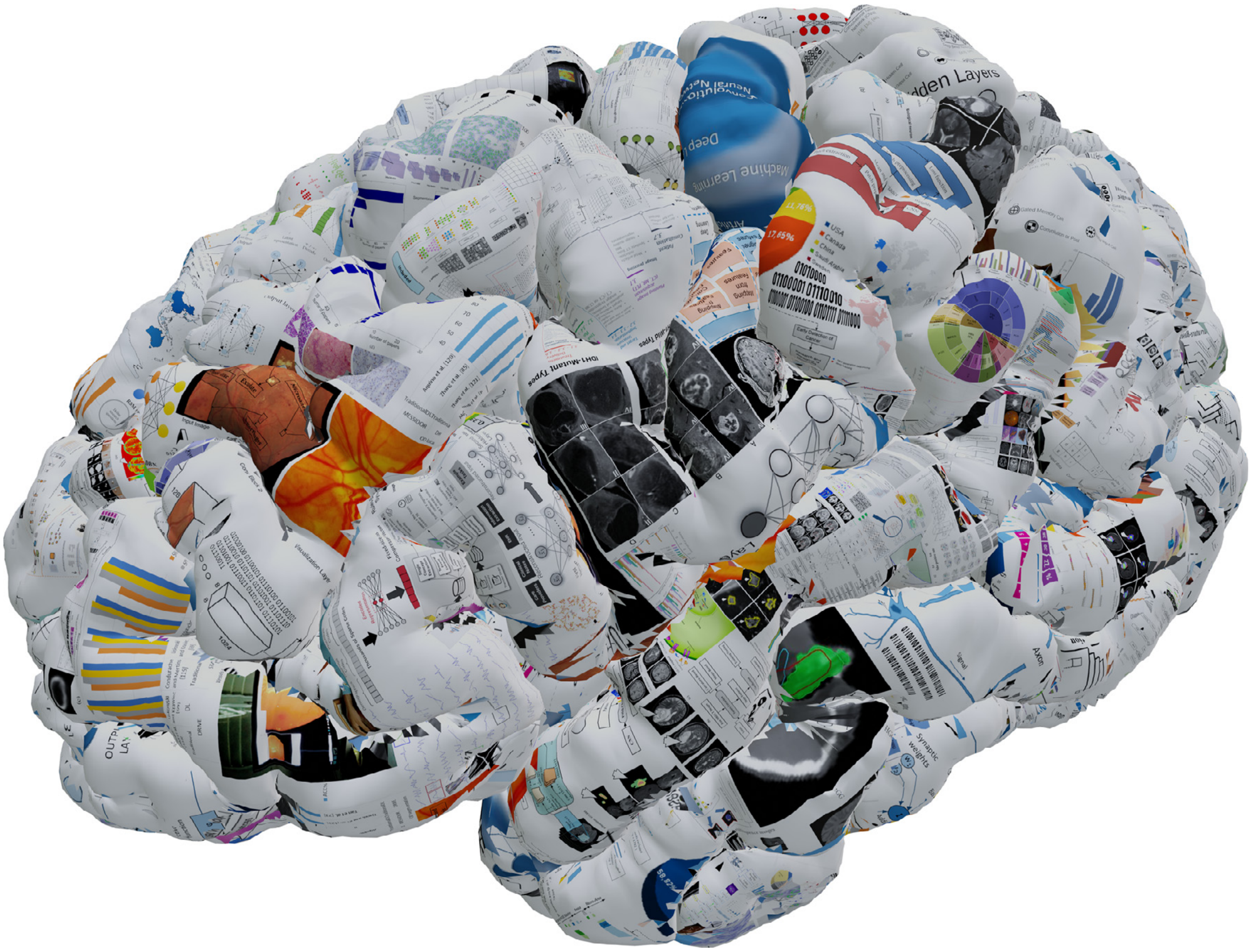

**Fig. 1.** Collage mapping all figures from the reviewed articles to a left hemisphere brain surface.

### 2.1. Medical deep learning reviews in 2017

With the described search strategy, seven medical deep learning surveys published in 2017 were discovered. Fig. 3 shows a network visualization of the review article keywords from 2017 revealing the keyword “humans” with its connections as the main cluster. Further main keyword clusters are “deep learning” and “neural networks (computer)”, which also reveal the main commonalities and trends for the surveys in 2017. More specific topics in the surveys of 2017 are “electronic health records” and “diagnostic imaging”. The reaming clusters are of a more general nature, like “machine learning”, “algorithms” and “image processing”. The presented reviews from 2017 cite 1060 contributions and have been cited 6089 times (status as of August 2020). They cover the following categories and are ordered by *epub* (electronic publication) date in 2017 (see Table 2):

- Medical image analysis (I.);
- Healthcare;
- Medical image analysis (II.);
- Stroke management;
- Analysis of molecular images in cancer;
- Health-record analysis;
- Microscopy image analysis.

*Medical image analysis (I.)* – The aim of medical image analysis is to automatically or semi-automatically extract information from patient data. For instance, this could be an automatic determination of the tumor volume from a patient’s magnetic resonance imaging (MRI) scan with the aim to choose the appropriate therapy strategy. Shen et al. [28] introduce in their publication the basics of deep learning-based approaches and survey their suc-cess in fields like image registration, tissue segmentation, anatomical/cell structures detection, computer-aided disease diagnosis, but also computer-aided disease prognosis. They conclude their work by pointing out remaining research challenges and give suggestions for future research directions that could advance medical image analysis.

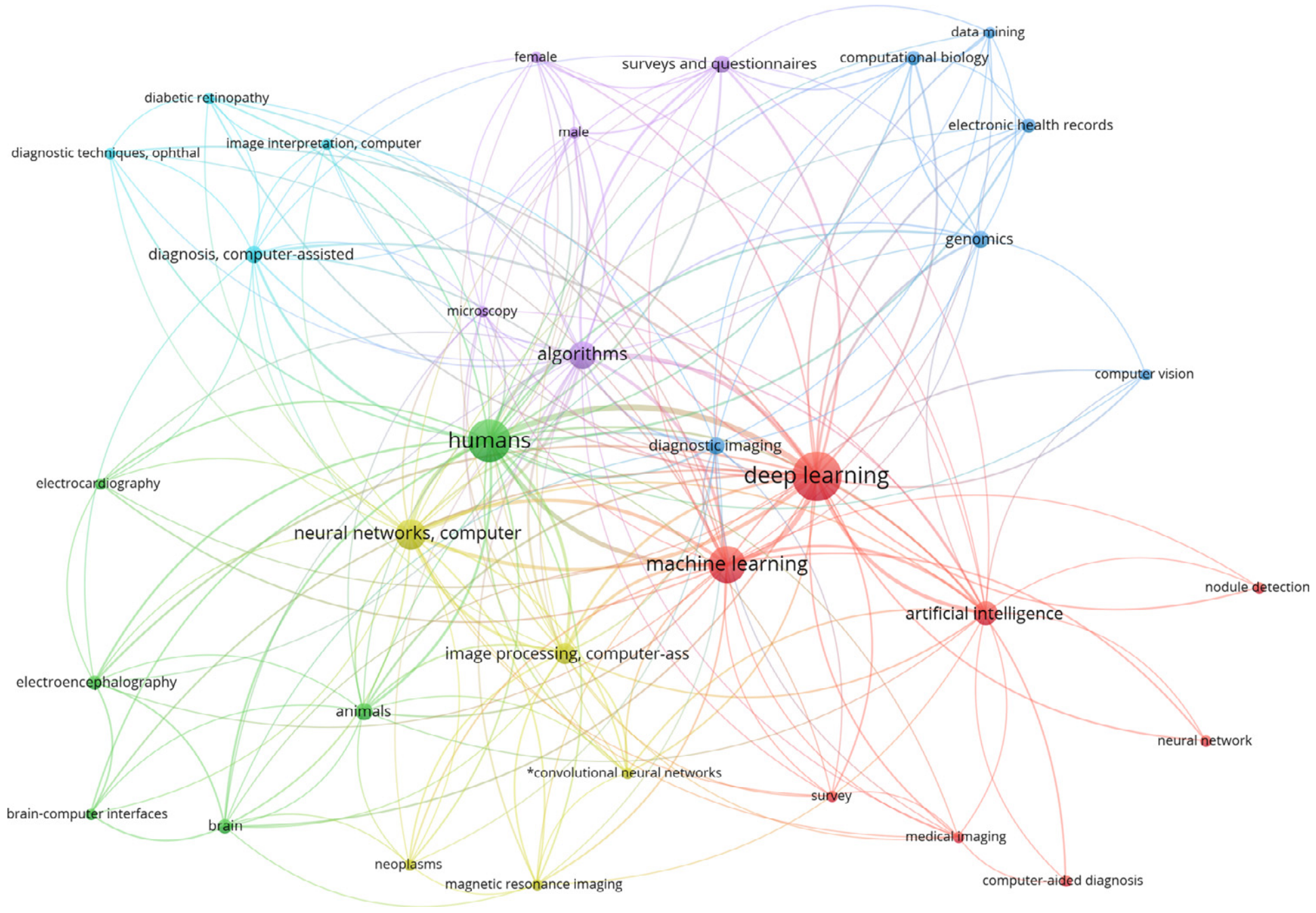

**Fig. 2.** Network visualization for the review articles supplied keywords from 2017 to 2019 performed with VOSviewer.

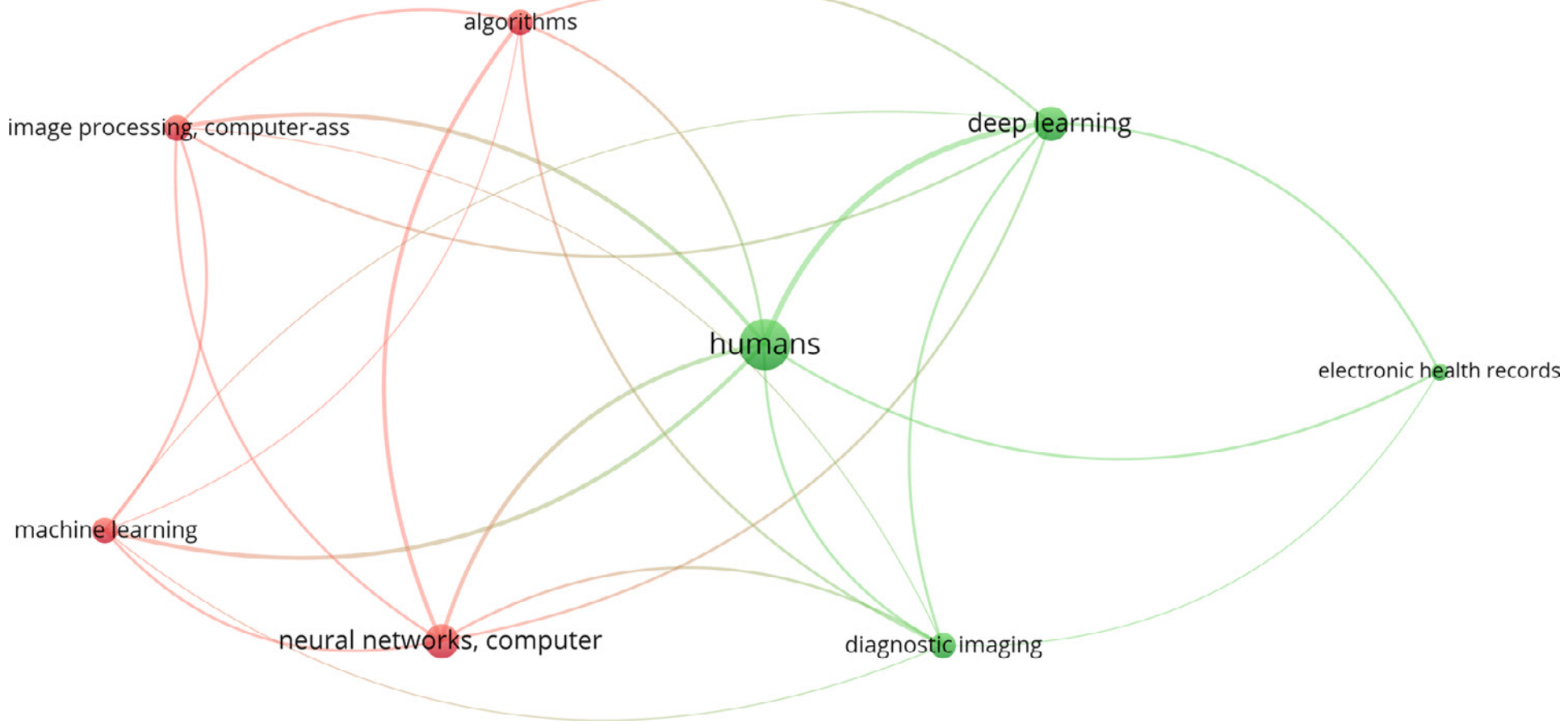

**Fig. 3.** Network visualization for the review articles supplied keywords in 2017 performed with VOSviewer.

*Healthcare* – The umbrella term healthcare envelopes the maintenance and advancement of people's health by diagnosis, prevention, treatment, but also recovery or even cure of illness, disease, injury, or any further physical or mental maladies. In that context, the survey article of Miotto et al. [29], reviews published research using deep learning-based approaches and technologies to improve the healthcare field. Centered on the analyzed publications, they conclude and propose that deep learning-based methods can be used to advance human health by exploring and exploiting big biomedical data. Furthermore, they depict limitations and the need for improved methods and applications and discuss future challenges in this area.

*Medical image analysis (II.)* – The publication of Litjens et al. [30] surveys the main deep learning-based concepts that are relevant for the area of medical image analysis. They summarize over 300 works within the area and analyze the usage of deep learning-based methods for object detection, image classification, segmentation, but also registration and further tasks. Moreover, they give compact, categorized outlines of studies in different areas of application, namely digital pathology, neurological, pulmonary, retinal, breast, as well as abdominal, cardiac, and musculoskeletal imaging. Finally, they give a summary of the current works at that time and discuss the remaining research questions and directions for upcoming research contributions.

*Stroke management* – Stroke can cause a long-term disability and a vast amount of research has been focused on using neuroimaging to explore regions of ischemia, which have not been affected by cellular death. In this context, Feng et al. [31] review clinical applications for deep learning-guided stroke management. They identify the following core topics for translating deep learning-based methods in the management of strokes, namely image segmentation, multimodal prognostication, but also radiomics (automated featurization).

*Analysis of molecular images in cancer* – Molecular imaging is of major interest for early cancer detection, because it opens the possibility to visualize biological changes on a molecular, but also on a cellular level, which enables a quantitative analysis of them. Hence, Xue et al. [32] published a survey about deep learning-based applications for an automated analysis of molecular cancer image acquisitions. They survey the deep learning-based applications in the field of molecular imaging with regards to a segmentation of tumor lesions, classification of tumors, and a prediction of patient survival.

*Health-record analysis* – Health-record analysis explores the digital information stored in electronic health databases. The initial intention for storing information of patients are administrative tasks in healthcare, such as billing. However, subsequently health records also became interesting for numerous applications in clinical informatics for researchers. Hence, Shickel et al. [33] perform a review about deep learning-based research for clinical applications that depend on the analysis of health-record data. They explore numerous deep learning-based frameworks and techniques that have been used for various clinical tasks; for example, information extraction, representation learning, outcome prediction, phenotyping, and de-identification. The authors discovered several remaining research challenges, such as heterogeneity of data, the lack of available universal benchmark tests, and the interpretability of models. They finalize their analysis by recapitulating the recent works, as well as pointing out directions that could be upcoming research topics in deep learning-based processing of health-records.

*Microscopy image analysis* – Microscopy images are images acquired from a microscope that can be utilized for the characterization of various diseases, such as brain tumors, breast cancer or lung cancer. Xing et al. [34] explore the image analysis domain for medical microscopy by providing at first a dense overview of common deep neural networks. Then, they analyze and review state-of-the-art results of deep learning in the analysis of microscopy images, for example in the tasks of image segmentation, object detection and classification. The authors also describe several architectures in deep learning, namely convolutional and fully convolutional neural networks, but also deep belief and recurrent neural networks, and lastly, stacked autoencoders. Thereby, they investigate and depict the specific network structures for the different applications in the analysis of microscopy images. The authors end their review by outlining remaining research needs, and by highlighting possible research directions in the domain of deep learning-based processing of microscopy images.

#### 2.1.1. Diving deeper: architectures, evaluations, pros, cons, challenges and future directions in 2017

Table 5 presents more details about the presented methods, pros, cons, evaluations, challenges and future directions for the reviews from the year 2017. All reported surveys share a number of important conclusions. They agree that deep learning is a promising approach for a wide variety of medical fields and tasks and predict that it will find increasing use in diagnosis, predictions, decision making and task automation. The deep learning-based methods explored by the respective surveys typically outperform previous state-of-the-art algorithms based on more naive approaches. In addition, the authors of the surveys all share the opinion that several challenges remain unsolved so far and will require additional exploration in the future. Among those are the inherently low explainability of deep learning approaches (often termed the "black box" problem) and lack of structured and expert-labeled or -annotated data, suggesting the creation of large-scale public datasets.

### 2.2. Medical deep learning reviews in 2018

With the proposed search strategy, 15 surveys were identified in medical deep learning from 2018. Fig. 4 shows a network visualization for the review articles supplied keywords in 2018 that reveals, equivalent to the surveys from 2017, the keywords "humans" and "deep learning", and its connections, as the main clusters. Further main keyword clusters center around the more general keywords "machine learning", "neural networks (computer)" and "algorithms". However, the smaller clusters around the keywords "electrocardiography", "computational biology", "surveys and questionnaires", "genomics" and "animals", show that the works in medical deep learning broadened in 2018 compared to 2017. The proposed reviews from 2018 themselves refer to 1684 contributions and have been cited 947 times (status as of August 2020). They cover the following categories, ordered by epub date in 2018 (Table 3):

- Toxicity of chemicals;
- Pulmonary nodule diagnosis;
- Physiological signals;
- DNA sequencing;
- Radiotherapy;
- Ophthalmology;
- Electronic health records;
- Bioinformatics;
- Personalized medicine;
- 1-D biosignals;
- Omics;
- Sport-specific movement recognition;
- Diabetic retinopathy;
- Image cytometry;
- Radiology.

*Toxicity of chemicals* – Toxicity testing and evaluation of chemicals is important for humans and animals, because they are exposed lifelong to natural and synthetic chemicals. Tang et al. [35] analyze in their work how deep learning-based tools can be a utilized for toxicity prediction, by building models for quantitative structure-activity relationships. They focus on large datasets, where classic data analysis techniques cannot deliver fast results. First, a technical overview about deep neural networks is provided by the authors. Then, recent works for the prediction of chemical toxicity models based on deep neural network approaches are explored. Finally, the important data sources for toxicity are outlined, remaining challenges are highlighted, and future directions for deep neural network-based approaches for the prediction of chemical toxicity are provided.

*Pulmonary nodule diagnosis* – A pulmonary nodule is a small, rounded opacity within the pulmonary interstitium. In their review, Yang et al. [36] present deep learning works that aid the decision-making in pulmonary nodule diagnosis. The deep learning-based methods they survey focus on computer-assisted feature extraction, false-positive reduction and nodule detection, but also on a benign-malignant classification in large volume scans of the chest.

*Physiological signals* – Physiological signals are signals from psycho-physiological measurements. In their survey, Faust et al. [37] review deep learning-based approaches utilized in healthcare applications that exploit physiological signals. Their bibliometric review revealed that the analyzed contributions focused mainly on Electromyograms (EMGs), Electroencephalograms (EEGs), Electrocardiograms (ECGs) and Electrooculograms (EOGs). Hence, they used these four categories to structure the content of their survey.

*DNA sequencing* – Deoxyribonucleic acid (DNA) sequencing is the determination procedure to reveal the order of nucleotides in DNA. Celesti et al. [38] review deep learning-based approaches to accelerate the process of DNA sequencing, given that huge amount of genomics data is emerging from next-generation sequencing (NGS) techniques. They provide a taxonomic analysis, by outlining the main deep learning-based NGS tools and software, and discuss

**Table 5**
Methods, pros, cons, challenges and future directions in medical deep learning in 2017.

| Publication | Methods | Pros | Cons | Challenges | Future Directions |
|---|---|---|---|---|---|
| Medical image analysis I* Shen et al. [28] | CNN, DBM, DBN, SAE | -DL can learn features through labeled data itself -Can be used experts outside of the medical domain | -Overfitting due to limited training samples | -Image features learnt by deep learning are difficult to understand and interpret | -Build medical equivalent of ImageNET -Incorporate domain-specific knowledge in design/training -Develop a universal algorithm compatible with various imaging modalities and protocols |
| Healthcare Miotto et al. [29] | AE, CNN, RBM, RNN | -DL can model, represent and learn from heterogenous EHR | -Neural networks need improvement in interpretability, data integration, and security | -Low data volume -Data heterogeneity -Low interpretability -Domain complexity -Disease temporality | -Use of federated learning, explainable AI -Modeling temporality -Include expert knowledge into modeling -Preserve privacy -upscale and standardize EHR |
| Medical image analysis II* Litjens et al. [30] | AE, CNN, DBN, GAN, RBM, SAE, VAE | -End to end training (CNN) -Freely available pre-trained deep learning models | -Hyper-parameter tuning is empirical -Subjective medical image annotation is susceptible variability and uncertainty | -Medical image annotation is time consuming and expensive | -Task-specific pre-processing and data augmentation techniques -Incorporate prior knowledge of the specific domain into training -Radiological reports could be used to annotate medical images -Leverage non-expert annotation through crowd-sourcing -Unsupervised learning using unlabeled data -Interpretable DL |
| Stroke Management Feng et al. [31] | CNN, DNN | -Can apply automated featurization, image segmentation, multi-model prognostication, CAD | -Neural networks need improvement | -DL requires substantial programming skills -Data scarcity | -DL will increasingly become a personalized medicine tool for stroke specialists due to its speed, power and versatility |
| Analysis of molecular images in cancer Xue et al. [32] | AE, CNN, DNN, SAE | -Improved speed and performance in tumor segmentation, classification, and survival prediction | -CNN may overfit -CNNs have time consuming training, challenging with low data | -Insufficient and imbalanced datasets -Subjective model depth, architecture and hyperparameters -Abstract high-level features | -Self-supervised approaches can solve the annotation problem and make larger datasets usable -Explore model optimization and explainability -Establish larger-scale public datasets |
| Health Record Analysis Shickel et al. [33] | AE, CNN. MLP, RBM, RNN | -LSTM, RNNs, and variant can process sequential data | -Lack of transparency and interpretability | -Heterogenous data -Lack of reproducibility and universal benchmarks | -Include robust mechanisms to handle EHR irregularity -Focus NLP on the clinical notes -Unify the representation of various types of patients' data -Patient deidentification using DL -Increase interpretability |
| Microscopy image analysis Xing et al. [34] | CNN, FCN, RNN, SAE | -Unsupervised training (SAE) -Unfixed input size (FCN) -easily parallelized training (CNN) | -Obtaining large number of annotated microscopy images is expensive -NN requires a fixed input size | -Low interpretability -Processing high volumes of medical data require computational acceleration | -Develop DL methods for WSI analysis -Use a patch-based strategy to reduce computational expenses -Fusing different types of patients' data -Design task-specific DL architecture based on domain knowledge -Develop unsupervised or semi-supervised learning algorithms |

Abbreviations: AE: auto-encoder, CAD: computer-assisted diagnosis, CNN: convolutional neural network, DBM: deep Boltzmann machine, DBN: deep belief network, DL: deep learning, DNN: deep neural network, EHR: electronic health record, FCN: fully convolutional network, GAN: generative adversarial network, MLP: multilayer perceptron, NLP: natural language processing, LSTM: long short-term memory, RBM: restricted Boltzmann machine, RNN: recurrent neural network, SAE: stacked auto-encoder, VAE: variational auto-encoder, WSI: whole slide imaging. *Also discussed in [74].

remaining research questions with a special focus on cloud computing.

*Radiotherapy* – Radiotherapy (or radiation therapy) utilizes ionizing radiation to control or kill malignant cancer cells. Therefore, treatment planning and delivery is complex and may be facilitated and partially automated by artificial intelligence. In their review, Meyer et al. [39] start explaining the fundamentals of deep learning-based techniques by relating them to the wider machine learning field. They give an overview of main network architectures, with special attention to convolutional neural networks. Afterwards, they analyze and summarize deep learning-based works for radiotherapy applications by classifying them into seven unique categories that are related to the workflow of the patient.

*Ophthalmology* – The diagnosis and treatment of eye disorders in medicine is called ophthalmology. In their review, Grewal et al. [40] explore deep learning as a new technology for ophthalmology with various possible applications. They explore deep learning-based methods that have been utilized in various diagnostic modalities, such as digital photographs, visual fields, and optical coherence tomography. They identify applications in the evaluation of numerous diseases, like cataracts, age-related macular degeneration, glaucoma, and diabetic retinopathy.

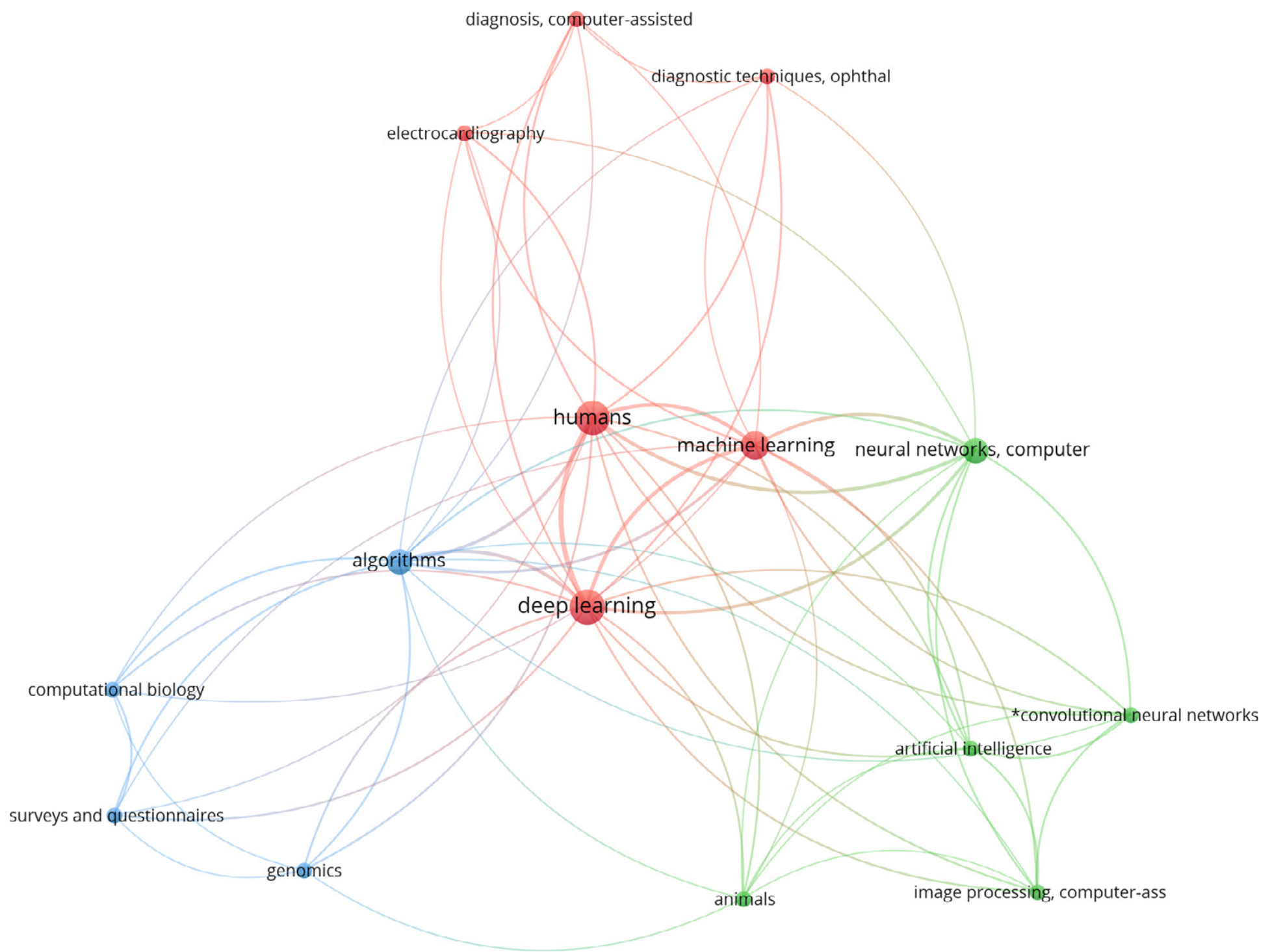

**Fig. 4.** Network visualization for the review articles supplied keywords in 2018 performed with VOSviewer.

*Electronic health records* – Electronic health records (EHR) summarize the data that is routinely collected from millions of patients across diverse healthcare centers, including information about patient demographics, diagnoses, medication prescriptions, clinical notes, laboratory test results, and medical images. Xiao et al. [41] performed a systematic analysis of deep learning-based models for exploring such EHR data by outlining them in regards to the kind of analytics task they perform and the kind of deep learning-based model architecture they use. They also depict the specific challenges resulting from such health data and tasks, and discuss potential solutions, as well as strategies for an evaluation in this field.

*Bioinformatics* – Bioinformatics is an interdisciplinary field developing approaches and software tools for the understanding of biological data with a strong focus on large and complex datasets. Lan et al. [42] survey research works combining deep learning-based methods with data mining, aiming to explore particular knowledge of the bioinformatics domain. The survey work gives a summary of several conventional algorithms in the data mining field that have been utilized for different tasks, like pre-processing, clustering and classification, but also of optimized neural network-based architectures and deep learning-based approaches. Finally, they outline the advantages and disadvantages in practical applications and discuss and compare them in terms of their industrial usage.

*Personalized medicine* – The aim of personalized medicine is to provide tailored patient-specific medical treatments via the identification of common features, like their inheritance, genetics and so on. Zhang et al. [43] provide a research outline concerning learning algorithms and methods, and their application, with an emphasis on deep learning-based approaches for personalized medicine. They explore three main application domains by giving insights into their pros and cons, namely disease characteristic identification, drug development, and a prediction of the therapeutic effect. They conclude that the analyzed learning algorithms and methods cannot be seen as a general solution for all kinds of medical problems.

*1-D biosignals* – Biosignals are electrical, thermal, mechanical or other signals measured over time, coming from the human body or other organic tissues, for example an ECG measures electrical activity originating from the heart muscle. Ganapathy et al. [44] survey deep learning approaches for 1-D biosignals in the field of computer-aided diagnosis. Further, they aim to establish a taxonomy to categorize the increasing number of applications in that area. The deep learning-based models were arranged according to the origin, type and dimension of the biosignal, the application goal, type and size of the ground truth data, type and schedule of network learning, and the overall model topology.

*Omics* – The emergence of big data has also involved the field of omics, including genomics, transcriptomics and proteomics. Zhang et al. [45] aim to give an entry-level overview, to understand the usage of deep learning approaches and methods for tackling problems and challenges in the omics domain. They outline and discuss various deep learning-based techniques that have fused deep learning with omics. Furthermore, they explore deep learning-based open-source frameworks with regard to their performances and features, but also highlight upcoming challenges and chances.

*Sport-specific movement recognition* – Sport-specific movement recognition can be utilized for the objective performance analysis of an (elite) athlete. In that regard, Cust et al. [46] explore the automated recognition and characterization of movements in sports, which can provide an alternative for an otherwise manual, time-consuming, limited performance analysis. The authors perform a systematical literature analysis on machine learning- and deep learning-based approaches for movement recognition in sports depending on input data from computer vision and inertial measurement units. They conclude that the experiment set-up, data pre-processing, and method development need to be considered and adjusted in accordance with the specific characteristics of the examined (sport) movements to achieve good results.

*Diabetic retinopathy* – Diabetes is on the rise worldwide and the most frequent microvascular complication is diabetic retinopathy, which can lead to visual impairment or even blindness. Nielsen et al. [47] performed a systematic review of deep learning techniques used for diabetic retinopathy screening. They explore works utilizing deep learning-based approaches for the classification of full-scale diabetic retinopathy, using retinal fundus images from diabetes patients. However, they only include works, which used a grading scale for diabetic retinopathy, a deep learning performance score, and have been compared to a reference standard from a human grader.

*Image cytometry* – Cytometry is the measurement of cell characteristics, like the cell size, cell count, cell morphology, cell cycle phase and DNA content. Gupta et al. [48] review how deep learning has been used to analyze microscopy image data of tissue samples and cells. They begin with an overview of neural networks and deep learning. They outline requirements for the input data, computational resources, and limitations and challenges in published works on deep learning in image cytometry, as well as identify methods that have not yet been used for cytometry data for potential future work.

*Radiology* – Radiology is the medical field of extracting useful information from images, like computed tomography (CT) or MRI, for diagnosis and treatment of humans and animals. In their review, Mazurowski et al. [49] give an introduction about the field of radiology and outline open research questions that could be tackled with deep learning techniques. They further provide an overview of basic deep learning concepts, such as convolutional neural networks. Next, they outline deep learning-based research contributions published within the radiology discipline. Thereby, they organize the reviewed works by the specific type of tasks they aim to support. They conclude their work by discussing remaining problems, but also highlight opportunities for using deep learning-based approaches within the practice of radiology.

### 2.2.1. Diving deeper: architectures, evaluations, pros, cons, challenges and future directions in 2018

Table 6 presents more details about the presented methods, pros, cons, evaluations and challenges and future directions for the reviews from the year 2018. Again, all the reported reviews share several important conclusions. Deep learning methods outperform machine learning methods over a wide variety of subjects, tasks, and datasets. All reviews predict an increase in (and increasing importance) of deep learning-assisted research and, at some future junction, practical applications. Most deep learning methods covered in the individual papers were CNNs and the review authors specifically cite CNNs as yielding impressive automatically extracted features and performances. The same general issues that were already reported in 2017, such as lack of interpretability and high-quality dataset availability, are reported again. Lastly, while generally considered promising, deep learning methods at this point have not been integrated into practical workflows.

## 2.3. Medical deep learning reviews in 2019

With the proposed search strategy, 21 surveys were identified in the area of medical deep learning in 2019. Fig. 5 shows a network visualization for the review articles supplied keywords in 2019 that reveals the keyword "deep learning" and its connections as the main cluster. Further main keyword clusters are "humans", "machine learning", and "artificial intelligence". New clusters arise around the keywords "brain" and "brain-computer interfaces", which shows that this organ has been heavily targeted by the research community in 2019. Also interesting is the cluster around "convolutional neural network", which shows that CNN gained momentum in the medical domain by 2019. The proposed reviews from 2019 refer to 2279 contributions and have already been cited 408 times (status as of August 2020). They are ordered by *epub* date in 2019 (Table 4) and cover the following categories:

- Medical imaging;
- Brain cancer classification;
- Electroencephalogram;
- Pulmonary nodule detection;
- Neuro-oncology;
- Diabetic retinopathy;
- Cardiac arrhythmia;
- Protein structure;
- Electroencephalography;
- Neurology;
- Cancer diagnosis;
- Ultrasound;
- Radiation oncology;
- Drug-drug interaction;
- Urology;
- Sleep apnea;
- Ophthalmic diagnosis;
- Alzheimer's disease;
- Pulmonary nodule detection;
- Liver masses;
- Pulmonary medical imaging.

*Medical imaging* – Medical imaging covers the field of producing visual representations of the internal body, for example using computed tomography, magnetic resonance imaging or ultrasound, just to name a few. Biswas et al. [50] explore various types of deep learning systems available, with a focus on current deep learning-based applications in medical imaging. They also outline the transition of technology from machine learning to deep learning and provide a complexity analysis and potential advantages for developers and users.

*Brain cancer classification* – In general, brain tumors are classified into several types, depending on whether they are, for example, benign or malignant, which helps to choose an optimal treatment for the patient. Tandel et al. [51] review machine learning and deep learning-based methods in the field of brain cancer, with a focus on pathophysiology. They include a review of imaging modalities and automatic, computer assisted methods for the characterization of brain cancer. Moreover, they outline the analysis of connections between cancer in the brain and additional brain disorders, such as Alzheimer's disease, Wilson's disease, Parkinson's disease, stroke, leukoaraiosis, and further neurological disorders.

*Electroencephalogram* – In the field of neuroscience, EEG analysis is an important technique with applications not only in neuroscience, but also neural engineering, like brain-computer interfaces (BCIs). Craik et al. [52] perform a systematic review on deep learning applications for EEG classification, addressing several questions, including specifiying specific EEG tasks. They analyze the studies based on several categories, like preprocessing algorithms for EEG, the kind of input, and the type of deep neural network architecture. The deep learning tasks were divided into five groups, namely the mental workload, emotion recognition, seizure detection, motor imagery, event related potential detection, and sleep scoring. For every kind of task, they outline the specific formulation of the input, classifier recommendations, and other major important characteristics.

*Pulmonary nodule detection* – Pehrson et al. [53] systematically reviewed the deep learning or machine learning-based methods used for the automatic detection of pulmonary nodules using a common dataset, the Lung Image Database Consortium and Image Database Resource Initiative (LIDC-IDRI) database. They divide the works into two subcategories based on their overall architecture.

**Table 6**
Methods, pros, cons, challenges and future directions in medical deep learning in 2018.

| Publication | Methods | Pros | Cons | Challenges | Future directions |
|---|---|---|---|---|---|
| Toxicity of chemicals Tang et al. [35] | DNN | -DNNs outperform ML in prediction of epoxidation, quinone formation, metabolite reactivity, classification of toxicity effects, and chemical-target interaction prediction | -Overfitting in DNNs -DNN underperformance with small quantities of training data | -Imbalanced, inhomogeneous, small datasets -Necessity of long training times and large computational resources | -Creation and curation of larger, public datasets by combining datasets from published works, patents and the web |
| Pulmonary nodule diagnosis Yang et al. [36] | AE, CNN, DBN, MTANN, SDAE | -Inclusion of CAD, feature extraction and benign-malignant classification -CNNs outperform SVMs and handcrafted rule-based algorithms | -Large amounts of data are required for successful training | -To facilitate DL, large datasets must be set up using time-consuming and unreliable manual labeling | -DL for decision support in pulmonary nodule diagnosis and classification -Alleviate the burden of dataset labeling with reinforcement learning -Create public datasets similar to ImageNet -Multi-scale patches during training to bridge data gap |
| Physiological signals Faust et al. [37] | AE, CNN, DBN, DNN, KNN, LSTM, RBM, RNN, SDAE, SVM | -Eliminates tedious and error-prone manual feature selection -Successful applications include state predictions, classifications and signal decoding. | -Time consuming -Model architecture and hyperparameters decided without statistical evaluation -Failure to capture information in a generalizable way for chaotic signals | -Long training times -Need for large training sets | -Testing DL applications in practical settings |
| DNA sequencing Celesti et al. [38] | AE, CNN, DNN, HMM, MLFF, RNN | -Integrated into software for gene expression analysis, genome analysis, SNP research, and early cancer detection -Computational efficiency and best performance/generalization | -Not discussed | -Most existing NGS library preparation devices, sequencing instruments, and software tools have not been designed to work in a clinical networked environment | -DL for comparative genomics, forensic biology, biological systematic field, virology) -Cloud computing services will provide scalability and data sharing possibilities |
| Radiotherapy* Meyer et al. [39] | AE, CNN, DNN, RNN | -Availability of large amount of training data -Increasing power of GPUs | -DL theories are empirically and experimentally obtained -Small noise, imperceptible to humans, could alter the output completely | -Building coherent, large and balanced medical datasets that represent real-world scenarios -Difficulty of interpretation | -Not discussed |
| Ophthalmology Grewal et al. [40] | CNN, others unnamed | -DL has superior performance compared to older automated methods -Successful application for early diagnosis of age-related macular degeneration, diabetic retinopathy, glaucoma | -Difficulty conveying quantitative results (such as disease severity) -Overfitting on uncorrelated features, noise, or dataset-inherent biases | -Overinterpreting results from neural networks -Variability in dataset labels, and medical definitions | -Retinal photography with smartphones and DL deep learning could enable self-ophthalmology and diagnoses -Integrate DL in the ophthalmologic routine |
| Electronic health records Xiao et al. [41] | AE, CNN, GAN, GRU, LSTM, RNN, UE | -DL: better performance and less manual feature engineering required -Availability of large and complex datasets in healthcare for training -Successfully applied to clinical event prediction, disease classification, phenotyping, text labeling, generating continuous medical time series | -Lack of interpretability | -Temporality and irregularity of EHR data with lack of labels and multi-modality -Lack of generalization | -Interpretable and transparent model creation and data curation |
| Bioinformatics* Lan et al. [42] | CNN, DBN, decision tree, DNN clustering, NB, KNN, RNN, SAE, SVM | -DL can learn knowledge from massive amount of data automatically | -DL requires large datasets for training -Dependent on high-end hardware -Lack of interpretability | -Data imbalance is prevalent in the medical domain | -Aggregate different ML algorithms -Fuse data from different modalities -Develop semi-supervised and reinforcement learning algorithms |
| Personalized medicine Zhang et al. [43] | ANN, Bayesian networks, CNN, DBN, DNN, linear regression, MLP, RF, SDAE, SVM | -More modern DNNs and CNNs outperformed older algorithms -Scale more efficiently with increasing dataset complexity -Feature recognition and structural association in structured data -Successfully applied for drug development, disease characteristics and therapeutic effects | -Have not been applied to large scale datasets -Human intervention is required to extract new knowledge and for safe action | -Dataset limited availability, uncertainty, idiosyncrasy, size -Lack of reproducibility overfitting, computational complexity -Data privacy, lack of clinical approval, intellectual property rights, genetic correlation validation | -Upgrade clinical data and integration of already developed algorithms -Develop more reliable automated feature selection -Field growth |

(continued on next page)

**Table 6** (*continued*)

| Publication | Methods | Pros | Cons | Challenges | Future directions |
|---|---|---|---|---|---|
| 1-D biosignals Ganapathy et al. [44] | AE, ANN, CNN, DBN, DNN, RBM, RNN | -Non-linearity and complexity handled well<br>-Good performance even with multi-modal or complex data<br>-Successfully applied to enhancement, detection, clustering, diagnostics, and prediction. | -Weaknesses not explicitly covered, only the inherent challenges | -Small and complex datasets, device specificity, noise<br>-Real-time requirements for clinical applications<br>-Missing ground truths | -Increase standardization of network topology and parameters |
| Omics Zhang et al. [45] | CNN, DBN, DNN, GRU, LSTM, MLP, RBM, RNN, SAE | -Successfully applied to DNA, RNA, protein structure analysis, gene expression regulation analysis, disease prediction, protein function analysis<br>-CNNs can analyze spatial information in images<br>-RNNs can analyze correlated features and time-series<br>-DNNs are highly adaptable to almost all types of data | -Older RNNs are unstable during training<br>-Data cleaning is time-consuming and labor-intensive<br>-More training data, computations resources, and higher data quality required<br>-Lack of interpretability | -Model selection and parameter tuning | -Increasing relevance of reinforcement learning, incremental learning, and transfer learning<br>-Mitigation techniques for the disadvantages of DL methods will continually be developed |
| Sport-specific movement recognition Cust et al. [46] | CNN, DTW, KNN, LSTM, MLP, HMM, NB, RF, SVM | -DL outperforms other ML methods in performance and computational efficiency<br>-Does not rely on heuristic features | -Not discussed | -Lack of uniformity in data acquisition | -Fusion of IMU and vision data in models |
| Diabetic Retinopathy Nielsen et al. [47] | CNN, DNN | -Reduced manpower due to automation, cost of screening, and issues relating to interrater reliability | -Lack of trust due to “black box” nature | -Risk of bias towards favorable results due to exclusion of difficult images from datasets<br>-Lack of interpretability | -Overcome challenges with prediction uncertainty, quality control and lack of interpretability |
| Image cytometry Gupta et al. [48] | AE, CNN, DNN, GAN, MLP, RNN | -Features are generated independently and automatically<br>-Use of “transfer learning”<br>-Successful application areas covered all modalities, tasks and scales | -Require large amounts of annotated data<br>-Lack of interpretability<br>-Overfitting and underfitting | -Requires computational resources and programming expertise<br>-Class imbalances can impede the generalization ability<br>-Lack of interpretability | -Combine hand-crafted features and neural network analysis for strong, grounded results |
| Radiology* Mazurowski et al. [49] | ANN, CNN | -Effective in medical image classification, segmentation, detection, reconstruction and registration | -DL only outperformed human experts in a minority of radiological tasks<br>-Introducing DL into clinical practice will cause legal and ethical issues | -Datasets are smaller and often imbalanced, leading to suboptimal training<br>-Proper clinical validation is often overlooked | -Optimally incorporate DL in existing radiology workflow |

Abbreviations: AE: auto-encoder, ANN: artificial neural network, CAD: computer-assisted diagnosis, CNN: convolutional neural network, DBN: deep belief network, DL: deep learning, DNN: deep neural network, GAN: generational adversarial networks, GPU: graphic processing unit, GRU: gated recurrent units, HMM: hidden Markov model, IMU: inertial measurement unit, KNN: K-nearest neighbors, LSTM: long short-term memory, ML: machine learning, MLFF: multi-layer feed forward, MLP: multi-layer perceptrons, MTANN: massive training artificial neural network, NB: Naïve Bayes, NGS: next-generation sequencing, RBM: restricted Boltzmann machine, RF: random forest, RNN: recurrent neural network, SDAE: stacked denoising auto-encoder, SNP: single nucleotide polymorphism, SVM: support vector machine,.
UE: unsupervised embedding, *Also discussed in [74].

They conclude that machine learning and deep learning methods can be used for the detection of lung nodules, even with a high level of sensitivity, specificity and accuracy, however, they also conclude that there is no general technique to evaluate the performance of machine learning methods and algorithms.

*Neuro-oncology* – Gliomas represent 80% of all primary malignant brain tumors. Shaver et al.'s [54] survey provides an overview of the recent deep learning-based approaches and applications utilized for glioma detection and outcome prediction. They focus on the pre-operative and post-operative segmentation of tumors, genetic tissue characterization, and further prognostication. They show and conclude that deep learning-based approaches and applications are promising research directions for the segmentation and characterization of gliomas, their grading, and for giving a survival prediction.

*Diabetic retinopathy* – Another survey about diabetic retinopathy was published by Asiri et al. [55]. They focus on deep learning-based computer-aided diagnosis (CAD) systems, which they structure into various stages such as lesion segmentation, lesion detection, and lesion classification of fundus images. Furthermore, they discuss pros and cons of published deep learning-based methods to accomplish these tasks.

*Cardiac arrhythmia* – Cardiac arrhythmias are most commonly detected by an ECG, mainly because of its low cost and convenient usage. For these reasons, every day, ECG data is acquired in large amounts in hospitals and homes, which, on the downside, prevents a detailed manual data inspection. Parvaneh et al. [56] perform a review of recent advancements on cardiac arrhythmia detection using deep learning. They outline existing works according to five different aspects, namely the used dataset, the input data type, the kind of application, the applied architecture model, and finally, the evaluation of performance. They conclude by presenting the shortcomings of the surveyed studies and discuss possible future upcoming research directions.

*Protein structure* – The three-dimensional form of local segments of proteins is called protein secondary structure. Wardah et al. [57] wrote a review on predicting the secondary structures of proteins with deep learning-based approaches such as neural networks. They start with a background section about the secondary structure of a protein and introduce the basics of artificial neural networks. The authors conclude that there are several challenges left for the *in silico* predictions of secondary protein structures.

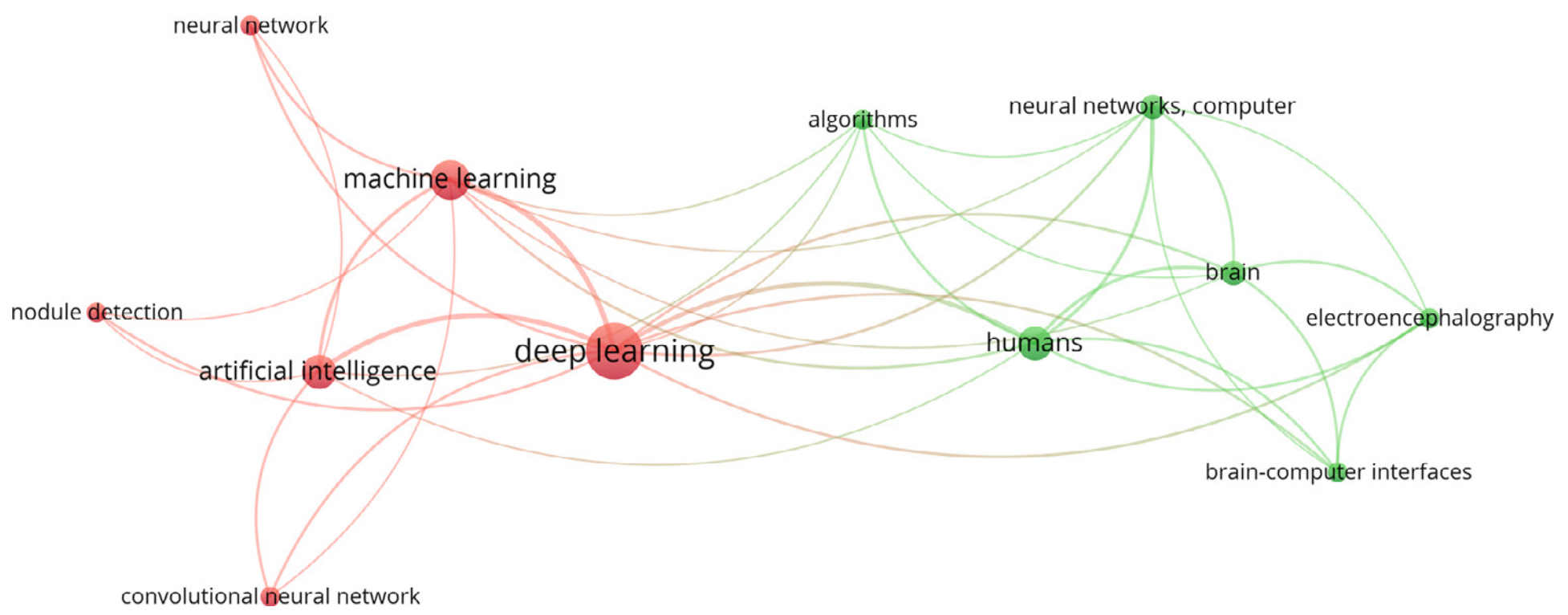

**Fig. 5.** Network visualization for the review articles supplied keywords in 2019 performed with VOSviewer.

*Electroencephalography* – As stated beforehand in the review about EEG classification, EEG analysis is an important yet difficult task, which requires several years of training because of its complexity. Roy et al. [58] performed a systematic survey of the analysis of electroencephalography with deep learning methods, covering various applications domains, like sleep, epilepsy, cognitive and affective monitoring, and brain-computer interfacing. In addition, they collected information pertaining to the data, such as the pre-processing methodology, the selection of the deep learning design, the results, and the experiments' reproducibility.

*Neurology* – The medical branch related to nervous system disorders (central and peripheral) is named neurology. Neurology covers the diagnosis and treatment of such disorders. Valliani et al. [59] review various neurology domains where deep learning algorithms have already been applied, like Alzheimer's disease diagnosis and early acute neurologic event detection. They also survey the segmentation of medical images for a quantitative evaluation of the neuroanatomy and vasculature structure, connectome mapping for Alzheimer's diagnosis, autism spectrum disorder (ASD), and attention deficit hyperactivity disorder (ADHD), as well as explore the granular genetic signatures and the signals of microscopic electroencephalograms.

*Cancer diagnosis* – A range of diseases, involving an abnormal growth of cells, which can also spread and invade other parts of the body, is called cancer. Munir et al. [60] give a bibliographic analysis on cancer diagnosis with deep learning-based approaches, starting with a background description of the cancer diagnosis domain. They cover the individual steps for cancer diagnosis, but also classification methods, like the asymmetry, border, color and diameter (ABCD) method, the Menzies method, the seven-point detection method, and pattern analysis. For each reviewed deep learning technique, they link to Python code. They also compile the applied deep learning models for different cancer types. Specifically, they discuss brain cancer, breast cancer, skin cancer and lung cancer.

*Ultrasound* – Ultrasound (US) is commonly used in the clinical routine due to it is nonionizing, low-cost, and portable characteristics, coupled with the ability of providing real-time images. Akkus et al. [61] present a review on deep learning-based applications in the ultrasound domain with the aim to improve the clinical workflow, including improving the acquisition of the US images, real-time evaluation image quality, objective detection and disease diagnosis, and in general, an overall optimized clinical workflow during ultrasound examinations. They also give a specific forecast of upcoming research trends and directions for deep learning-based methods that can facilitate an US diagnosis, but also reduce costs in health care, and provide an optimized clinical US workflow.

*Radiation oncology* – A physician or doctor who is specialized in the treatment of cancer using ionizing radiation, like radionuclides or megavoltage X-rays, is called a radiation oncologist. In that context, Boldrini et al. [62] perform a literature review in PubMed/Medline with a search strategy including the search terms "radiotherapy" and "deep learning". They identify recent publications on deep learning in radiation oncology, which they present with a focus on clinically oriented readers. The review shows how deep learning can support clinicians during their daily work, such as by reducing segmentation times, or predicting treatment outcomes and toxicities. However, they conclude that these techniques have yet to be employed in the clinical routine, and it remains to be seen how well they translate into practice.

*Drug-drug interaction* – Drug-drug interactions (DDIs) can cause adverse drug effects that have the potential to threaten public health and patient safety. Hence, these interactions are crucial for drug research and pharmacovigilance. Zhang et al. [63] review the state-of-the-art deep learning-based methods used for DDI exploration. They briefly outline every deep learning method from their surveyed studies and systematically evaluate their efficiency, strengths and weaknesses. They conclude their work by providing a discussion and giving an outlook on several future research challenges for the extraction of DDIs with deep learning-based approaches.

*Urology* – The medical branch of urology is focused on surgical and medical urinary-tract system diseases, including the urethra, urinary bladder, ureters, adrenal glands and kidneys. The urology branch also focuses on the reproductive organs of males, including the prostate, testes, penis, epididymis, seminal vesicles and vas deferens. Suarez-Ibarrola et al. [64] review recent and upcoming machine learning- and deep learning-based applications in the urology domain, with a focus on renal cell carcinomas, urolithiasis, prostate and bladder cancer. This covers, for example, the prediction of endourologic surgical outcomes in urolithiasis, the automatic distinction between malignant and benign small renal masses, the analysis of texture features and radiomics for the differentiation between low-grade and high-grade tumors in bladder cancer, MRI-based computer-aided diagnosis, biochemical recurrence prognosis, and prognosis algorithms for the Gleason score for prostate cancer.

*Sleep apnea* – Sleep apnea is a sleep disorders characterized by repeated stopping and starting of breathing. Sleep apnea can be scored with polysomnography, which is unfortunately expensive, inaccessible, uncomfortable and requires an expert technician. Mostafa et al. [65] preform a systematic review on the published deep learning-based research contributions used for detecting sleep apnea. They focus on exploring research subjects including the implementations of neural networks, a possible need for pre-processing or manual feature extraction, and finally, explore

**Table 7**
Methods, pros, cons, challenges and future directions in medical deep learning in 2019.

| Publication | Methods | Pros | Cons | Challenges | Future Directions |
|---|---|---|---|---|---|
| Medical Imaging Biswas et al. [50] | AE, (fully) CNN, DBN, DRN, FCN, SVM | -DBM has easy inference -Automated feature extraction -Learning of complicated and composite relationships in data -DL methods surpass in robustness and performance | -Unknown generalization capabilities (DBN) -Vanishing gradient problems during training (AE) | -Improvement needed before techniques could be integrated into clinical workflows -Only trained on small datasets | -Widespread use in research and clinical routine -Develop real-time applications |
| Brain cancer classification Tandel et al. [51] | ANN, CNN, EM, KNN, NB, RF, SVM | -Automatically produce features that are stable to deformation and translation invariant -DL outperforms other ML methods | -Computationally more expensive | -Not discussed | -Provide the fast, non-invasive diagnosis tool that the field needs |
| Electroencephalogram Craik et al. [52] | AE, CNN, DBN, LSTM, MLP, RBM, RNN, SAE, SVM | -Successfully applied to motor imagery, seizure detection, mental workload, sleep stage scoring, event related potential, and emotion recognition | -Not discussed | -Formulation of the input data (PSD, wavelet decomposition, etc.) | -Combine convolutions and recurrent or RBM architectures -Use de-noised EEG data |
| Neuro-oncology Shaver et al. [54] | ANN, CNN, CRNN, LSTM, SVM | -Do not require human-constructed features -CNN architectures provide high accuracies on segmentation, characterization, grading and survival prediction tasks | -Requires large quantities of annotated data, necessitating medical expert knowledge and significant amounts of time -Overfitting | -Lack of large amounts of annotated data | -Undisruptive integration into workflows -Work with regulatory bodies who currently restrict the use of ML/DL in clinical practice |
| Diabetic retinopathy Asiri et al. [55] | AE, CNN, DBN, RNN | -Automatic discovery of relevant features -Ability to train and deliver solutions in an end-to-end manner -Successfully applied to vessel and optic disk segmentation, lesion detection and classification, diabetic retinopathy diagnosis | -Require large amounts of labeled data -Tendency to overfit -Convergence of DL methods not always guaranteed -Lack of interpretability -Class imbalance of datasets | -Lack of large-scale annotated uniform training data -Generalization of DL methods | -More standardization in data, labels, and test metrics -Research GANs |
| Cardiac arrhythmia Parvaneh et al. [56] | AE, CNN, DBN, LSTM, RNN | -Unsupervised information capture and feature generation | -Highest scoring ML outperformed best DL -Overfitting | -Lack of interpretability -Large datasets needed | -Research interpretability -Identify optimal dataset sizes for training and testing |
| Protein structure Wardah et al. [57] | ANN, CNN, GRU, HMM, RNN | -Automatic protein structure prediction -Reduced time and costs compared to traditional in vitro analysis | -Not discussed | -Need in vitro techniques to determine hard truths, limiting datasets -Lack of comparability | -Automated prediction methods will drive the benchmark in the field closer to the theoretical accuracy boundary (approx. 88%) |
| Electrocephalography Roy et al. [58] | AE, CNN, DBN, GAN, MLP, RBM, RNN, SDAE | -Avoids time-consuming traditional feature engineering and provides end-to-end solutions -Can flexibly work with either small or large amounts of data -Can generalize to other tasks or datasets -Successfully applied to tasks including brain–computer interfacing, sleep staging, epilepsy, cognitive and affective monitoring | -Lack of reproducibility and interpretability | -Lack of labeled data -Dataset augmentations and hyperparameter searches are difficult to identify | -Efforts in reproducibility -Exploratory research into data quantity vs performance |
| Neurology Valliani et al. [59] | AE, CNN, DNN, GAN, GRU, LSTM, NB, RNN, SVM | -No manual feature crafting -Performance gains with larger datasets -Successfully applied for medical image classification, segmentation, functional connectivity, classification of brain disorders and risk prognosis | -Require large amounts of data to learn -High quality labels are time-consuming to create -Overfitting -Lack of interpretability | -Medical data suffers from heterogeneity and complexity -Data privacy, accessibility and ethical concerns over potential biases | -Research into generalizability and interpretability |
| Cancer diagnosis Munir et al. [60] | AE, AFINN, (fully) CNN, DBN, GAN, LSTM, RBM, RNN | -Learn features from raw images instead of requiring manually constructed features -Successfully applied to cancer diagnosis on multiple image modalities | -Require large datasets, generally with labels, a major time/cost investment | -Lack of available datasets -Datasets suffer from a strong disparity between positive and negative samples | -Not discussed |
| Ultrasound Akkus et al. [61] | AE, (fully) CNN, RBM, RNN, SDAE, SVM | -DL outperformed ML in generalizability -Successfully applied to detection, classification, segmentation, and diagnosis of lesions and nodules | -Lack of interpretability and explainability | -Dataset quality and performance vary in acquisition and interpretability -Size and quantity of public datasets are limited | -Clinical workflow and cost can be reduced -Include 3D, multiview cine clips, or spatiotemporal data into AI models |

(*continued on next page*)

**Table 7** (*continued*)

| Publication | Methods | Pros | Cons | Challenges | Future Directions |
|---|---|---|---|---|---|
| Radiation Oncology Boldrini et al. [62] | ANN, (fully) CNN, DNN, GAN, SVM | -Can analyze unstructured data and extract non-linear features without human supervision<br>-Capable of dimensional reduction<br>-Successfully applied to segmentation, outcome, response, and survival predictions | -Loss of functions are non-convex and no algorithm can guarantee to find an optimal solution<br>-Overfitting | -Need for expert knowledge in oncology and DL for dataset curation and training | -Need for bigger standardized datasets |
| Drug-drug interaction Zhang et al. [63] | CNN, GRU, LSTM, RNN, recursive neural network | -No need for manual feature engineering<br>-CNNs can generate translation-invariant descriptions from data<br>-RNNs can selectively hold relevant information in memory and analyze arbitrary length text inputs | -Tendency to be unstable during training<br>-Lack of interpretability | -Unstructured data and class imbalances | -Semi-/self-supervised learning, joint learning models, N-ary relation extraction, feature enrichment, interpretable modeling |
| Urology Suarez-Ibarrola et al. [64] | ANN, CNN, SVM | -Details not discussed | -In some cases, ML/DL were favorable to human raters, but traditional statistical methods outperform them, particularly in the field of urolithiasis | -Equipment variants and non-standardized data collection<br>-Generalization<br>-Heterogeneity of employed models and datasets | -Create large-scale public datasets<br>-Keep downscaling in mind to employ DL methods in real-time or on mobile devices |
| Sleep Apnea Mostafa et al. [65] | CNN. DBN, GRU, LSTM, MLP, RNN, SSAE | -Increased performance of DL vs ML methods | -Details not discussed | -Imbalanced heterogenous datasets<br>-Hyperparameter search | -Not discussed |
| Ophthalmic diagnosis Sengupta et al. [66] | (fully) CNN, FNN, MBNN, RF, SSAE, SVM | -DL outperforms for lesion and vessel segmentation, acute macular degeneration, glaucoma and diabetic retinopathy classification | -Requires large amounts of annotated data for training<br>-Can suffer from domain shift between training and test sets<br>-Generalizability | -Class imbalance<br>-Data acquisition and performance indicators are heterogeneous across reported papers | -Research generative models to augment existing datasets or balance classes<br>-Domain adaptation |
| Alzheimer's disease Ebrahimighahnavieh et al. [67] | AE, CNN, DBN, DNN, DPN, HMM, DBM, RBM, SVM | -Suited for modeling non-linear relationships<br>-Robust against translation and transformations of target features<br>-Capable of automated feature generation | -Require large amounts of data for training<br>-Loss of generalization capability<br>-Overfitting, computational cost and robustness | -Unpublished code bases<br>-Dataset imbalances and lack of data<br>-ROI-based methods require extensive domain expert knowledge | -Public benchmarking platform for fair comparisons of models<br>-Explainable AI<br>-Generation methodology |
| Pulmonary nodule detection Li et al. [68] | (MT)ANN, CNN, SDAE | -MTANNs and SDAEs can learn with fewer training examples than CNNs and generate new data easily<br>-Successfully applied to detection and classification of pulmonary nodules | -Longer training times and greater dataset requirements<br>-Small datasets in medicine<br>-Overfitting | -Heterogeneity of results | -Research into consistent, standardized integration of DL into clinical workflow |
| Liver masses Azer [69] | (fully) CNN, GAN | -Successfully applied to detection, classification, and segmentation of liver masses | -Details not discussed | -Heterogeneity of results | -Standardize reporting, report multiple performance metrics, practically apply, reproduce<br>-Collaborative data acquisition<br>-Case control studies to compare DL methods with human raters |
| Pulmonary medical imaging Ma et al. [70] | ANN, (fully) CNN, DPN, neural hy-pernetwork | -Self-learning and generalization<br>-Can extract information both from simple and complex data structures | -High computational and dataset size requirements<br>-Lack of interpretability | -Class imbalances in datasets<br>-Varying image quality | -Make use of unlabeled medical data to ease the annotation bottleneck<br>-Develop more interpretable DL models |

Abbreviations: AE: auto-encoder, AFINN: adaptive fuzzy inference neural network, ANN: artificial neural network, CNN: convolutional neural network, CRNN: convolutional recurrent neural network, DBN: deep belief network, DBM: deep Boltzmann machine, DL: deep learning, DNN: deep neural network, DPN: dual path network, DRN: deep residual network, EM: expectation maximization, FCN: fully connected network, FNN: feed-forward neural network, GAN: generational adversarial networks, GRU: gated recurrent units, HMM: hidden Markov model, KNN: K-nearest neighbors, LSTM: long short-term memory, MBNN: Multi-branch neural network, ML: machine learning, MLP: multi-layer perceptrons, MTANN: massive training artificial neural network, NB: Naïve Bayes, RBM: restricted Boltzmann machine, RF: random forest, RNN: recurrent neural network, SAE: stacked auto-encoder, SSAE: Stacked sparse auto-encoder, SVM: support vector machine.

the reported applications in terms of implementation and performance. The applied sensors, signals, databases and implementation difficulties have also been taken into consideration for an automatic, deep learning-based scoring process.

*Ophthalmic diagnosis* – Sengupta et al. [66] provide another review on ophthalmology, focusing on ophthalmic diagnosis using deep learning approaches based on fundus images (the back surface of the eye). They discuss recent deep learning approaches for diabetic retinopathy, glaucoma and age-related macular degeneration, and describe numerous datasets consisting of retinal images, which can be processed for deep learning-based ophthalmic tasks. Areas of applications from their surveyed works include segmen-tation of the optic cup, optic disk, and blood vessels, as well as lesion detection.

*Alzheimer's disease* – In developed countries, Alzheimer's Disease (AD) is one of the leading causes of death. AD is a chronic neurodegenerative disease that often starts slowly, but progressively worsens in the long-term. In this regard, Ebrahimighahnavieh et al. [67] systematically reviewed deep learning-based methods for an automatic AD detection from neuroimaging. They focus on the extraction of effective features and biomarkers, like genetic data, personal information, and scans of the brain, as well as required pre-processing steps and tips for handling neuroimaging data that comes from single- or multi-modality investigations. Moreover, they compare the performance of the deep learning models in AD detection and discuss remaining challenges, including the applied training strategies and datasets that can be accessed.

*Pulmonary nodule detection* – Another systematic review on deep learning-based methods in pulmonary nodule detection was published by Li et al. [68]. They focus on the detection and classification of nodules using CT scans not from the LIDC-IDRI database. They found that three types of deep learning architectures are commonly used, namely convolutional neural networks, deep stacked denoising autoencoder extreme learning machine (SDAE-ELM) methods and massive training artificial neural networks (MTANN). They conclude that high accuracy, specificity and sensitivity scores can be obtained with deep learning-based approaches in nodule classification and detection using CT scans not from the LIDC-IDRI cases.

*Liver masses* – A liver mass is a lesion in the liver that can be caused by an abnormal cell growth, a cyst, hormonal changes, or an immune reaction, but is not necessarily cancer. Azer [69] performs a systematic analysis on deep learning-based approaches, specifically convolutional neural networks (CNNs), for the detection of liver masses as well as hepatocellular carcinomas (HCCs). PubMed, the Web of Science, EMBASE and further research books were searched systematically, thereby identifying works analyzing cellular images, pathological anatomy images, and radiological images of liver masses or HCCs. The level of accuracy and CNN performance in cancer detection were presented with a focus on analyzing the kinds of liver masses and cancers and determining the image types which proved optimal for the precise detection of cancer.

*Pulmonary medical imaging* – Ma et al. [70] present an analysis on deep learning-based approaches for pulmonary medical imaging. Topics include classification, detection, and segmentation tasks in regard to pulmonary medical images, but also benchmarks and datasets. They provide an outline of the reviewed approaches, which have been implemented for different diseases of the lung, such as pneumonia, pulmonary embolisms, pulmonary nodules, and interstitial lung disease (ILD). Finally, they discuss the future challenges and potential directions in the area of medical imaging with deep learning techniques.

#### *2.3.1. Diving deeper: architectures, evaluations, pros, cons, challenges and future directions in 2019*

Table 7 presents more details about the presented methods, pros, cons, evaluations and challenges and future directions for the reviews from the year 2019. Interestingly, while most reviews cite largely the same advantages and disadvantages for deep learning, authors occasionally disagree on whether specific aspects of neural networks pose advantages, disadvantages or challenges, particularly concerning data availability. Some studies have had success training with very small datasets, while others did not, suggesting that not all the nuances of data pre-processing, augmentation, and training processes are fully understood yet. Many reviews report that individual papers could not be fairly compared in terms of performance due to the heterogeneity of methods and key performance indicators used, as well as due to the manifold differences in both datasets and data acquisition between reported papers. There appears to be a significant research gap in terms of standardization for these issues. Sometimes simpler statistical methods or traditional machine learning outperform deep learning and occasionally deep learning is reported to work better when shallower architectures are used, but typically deep learning methods handily outperform any competitors except human raters with years of experience. CNN architectures are typically used/reported the most often in the various review papers and many authors specifically report that CNNs appear to dominate the field both in terms of performance and prevalence. Lastly, deep learning methods are not deployed in clinical practice despite regularly achieving state-of-the-art results. Authors typically cite ethical concerns due to lack of interpretability, potential lack of generalizability, and unknown (or unknowable) biases as the reason. Thus, practical applications and real-world performance testing of newly developed deep learning methods, as well as deeper investigations into Explainable AI, constitute significant research gaps.

## 3. Conclusion

In this work, reviews and surveys on medical deep learning are presented in a systematic meta-review contribution. A systematic search has been performed in the common medical search engine PubMed, which resulted in over 40 review or survey publications published during the last three years. In addition to a brief summary of each survey, the references and citations of these reviews are presented (status as of August 2020).

Before 2017, no medical deep learning review article has been indexed under PubMed according to the proposed search strategy. This is easily explainable, because even though these kind of approaches had already been suggested and applied at the end of the last century [71,72], deep learning-based approaches only started to gain massive popularity after the convolutional neural network architecture AlexNet [73] won the ImageNet challenge in 2012. From that moment on, deep learning and convolutional neural networks have received inexorably increasing attention in various communities, including medical image analysis. However, it took some time to have enough published works for the first review or survey articles. In addition, there is also a massive number of review and survey articles in other, general disciplines. To give a rough impression of these, we performed an additional non-systematic search, which is, however, far from complete and the results are only presented in a systematic listing, because these works would go far beyond the scope of this contribution. Nonetheless, they may be an inspiration for interested readers and we arranged them in three categories (more details can be found in [74]):

1. Computer vision

   Object detection [75–77]
   Image segmentation [78,79]
   Face recognition [80–82]
   Action/motion recognition [83,84]
   Biometric recognition [85,86]
   Image super-resolution [87]
   Image captioning [88]
   Data augmentation [89]
   Generative adversarial networks [90]

2. Language processing

   General language processing [91]
   Language generation and conversation [92–95]

Named entity recognition [96,97]
Sentiment analysis [98,99]
Text summarization [100]
Answer selection [101]
Word embedding [102,103]
Financial forecasting [104]

3. Further works

Big data [105–107]
Reinforcement learning [108–110]
Mobile and wireless networking [111]
Mobile multimedia [112]
Multimodal learning [113]
Remote sensing [114]
Graphs [115]
Anomaly detection [116]
Recommender systems [117]
Agriculture [118]
Multiple areas [119–121]

## 4. Discussion

Typically, new trends in image processing are applied at first to general computer vision tasks, for example to 2D photos, before they are adapted and translated to tasks in the medical domain. This has several reasons. Firstly, 2D image processing is much less computationally intensive compared to processing large 3D image volumes from CTs or MRIs. Secondly, the algorithms are in general more "complex" and sophisticated (in terms of implementations) for 3D volumes than for 2D image processing, because they need to process one or more dimensions (if several scans at different time points have been acquired). Thirdly, often several image modalities and volumes, like combined positron emission tomography-computed tomography (PET-CT) scans, are available, and processing them jointly leads to information gain, but also increases the complexity. This is even more cumbersome if scans from different time points and/or different modalities, like CT and MRI, are not registered to each other. Finally, yet importantly, medical data is much harder to acquire and collect than for example natural images, especially in large quantities, not only because of the very time-consuming, often slice-by-slice manual ground truth generation and memory capacities, but also because of privacy concerns. Medical data is usually highly sensitive and personal, and therefore, using it for research purposes requires institutional review board (IRB) approvals and patient consent. Generally, data has to be pseudonymized / anonymized, by removing meta-information from the images and corresponding files, including name, sex and birth date. However, this is relatively easy compared to patient information that is encoded within the images themselves, like the patient's face in a head scan. Removing the eye area for patient de-identification within a 3D volume is possible, but laborious, because it must be done manually for every scan to make sure the volumes are properly de-identified. A fully automatic approach is conceivable, yet highly risky and potentially disastrous if it fails for even a single case. For head scans, de-identification by removing the eye area can be an option if the research is performed on a structure in another area of the head, like the lower jawbone [122,123], but on the downside, it can render the images unusable for applications requiring the entire volume, for example, facial-based medical augmented reality for the head and neck regions, for which all facial features are needed [124,125]. The IRB may allow the usage of the medical data for research purposes, but only within their own institution. This means that researchers from other institutions cannot re-use the existing data to validate published results or build upon existing methods to push the research boundaries. As a result, a considerable amount of effort has to be invested into obtaining IRB approval, acquiring data, and de-identification, which may delay new research by a few months, at best. Therefore, for large collections of rare pathological cases, it can easily take several years to establish a comprehensive database. Nonetheless, and against all odds, the massive amount of medical deep learning contributions is still increasing, and the proposed search strategy already reveals around 50 reviews or surveys for deep learning in PubMed by August 2020, which is more than all reviews from 2017 to 2019 together (Fig. 6). Equivalent to [58], we also looked at the locations of the first author's affiliations to get a sense of the geographical distribution of the medical deep learning reviews from this meta-review, and it reveals that the hotspots are the USA and China (Fig. 7). In case the first author provided several affiliations, we chose the very first one listed in the article.

The large amount of survey and review papers on medical deep learning published within the last three to four years is an indicator of the massive influence and importance that these algorithms already have in the medical community, and resulting clinical applications. This meta-review shows that, on average, a medical deep learning review has been published almost every month during the last years, with an approximately exponential increasing trend that seems to continue, if the distribution into the year 2020 is considered. Another indicator for the impact of deep learning in the medical field is the number of references (>5.000) and citations (>7.000) of the reviewed works. Besides the successes in outperforming state-of-the-art methods, there are several further reasons for and increase in research activities in (medical) deep learning:

- (1) The relatively easy application of deep learning algorithms to new data, enabled by comprehensive and user-friendly libraries and toolkits, like TensorFlow [126], PyTorch [127] or Caffe [128], just to name a few. These frameworks do not necessarily require an in-depth education in computer science. In contrast, in the era before deep learning, very good coding skills in programming languages like C or C++ were required to implement complex image processing algorithms. Factors like optimization for a reasonable runtime played a much larger role, as hardware was much weaker a few years ago.
- (2) Related to the first reason, most deep learning libraries and toolkits support Python bindings, which is a high-level, interpreted programming language, and therefore, easier to learn, apply and deploy compared to the aforementioned, compiled programming languages, like C or C++.
- (3) Relating to hardware, the broader availability of graphical processing units (GPUs) certainly contributed to the large distribution and application of medical deep learning and deep learning in general. Pretty much all deep learning libraries and toolkits natively support optimized training and execution of algorithms on a GPU, which speeds up the computation time many times over and makes many interesting big data applications possible. High-capacity GPUs decreased in price over the last few years, and GPU clusters, nowadays usually available at universities, research centers and companies, enable further parallelization and faster processing. Furthermore, GPU cloud servers and services (e.g. from Google Cloud or Amazon Web Services) can be accessed by everyone.
- (4) Another reason for the rapid spreading and adoption of deep learning (note that there is also already a review about artificial intelligence / deep learning techniques in imaging data acquisition, segmentation and diagnosis for covid-19 [129], and another one is on the horizon [130]), is that many researchers make their code publicly available to the research community, which is easily possible thanks to online repositories, like GitHub or GitLab. Because most implementations use common

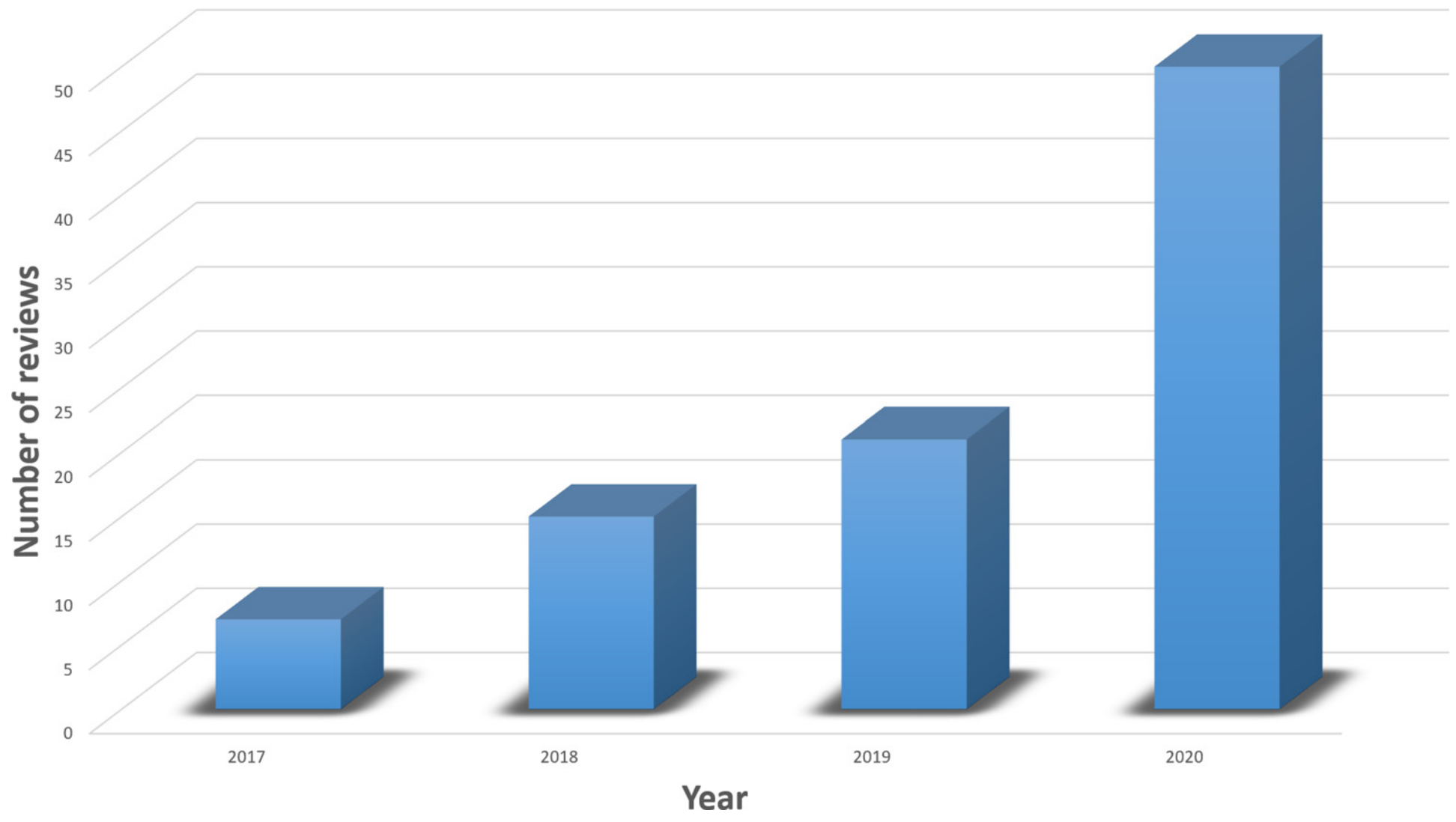

**Fig. 6.** Review and survey articles for medical deep learning in PubMed over the years (status as of August 2020).

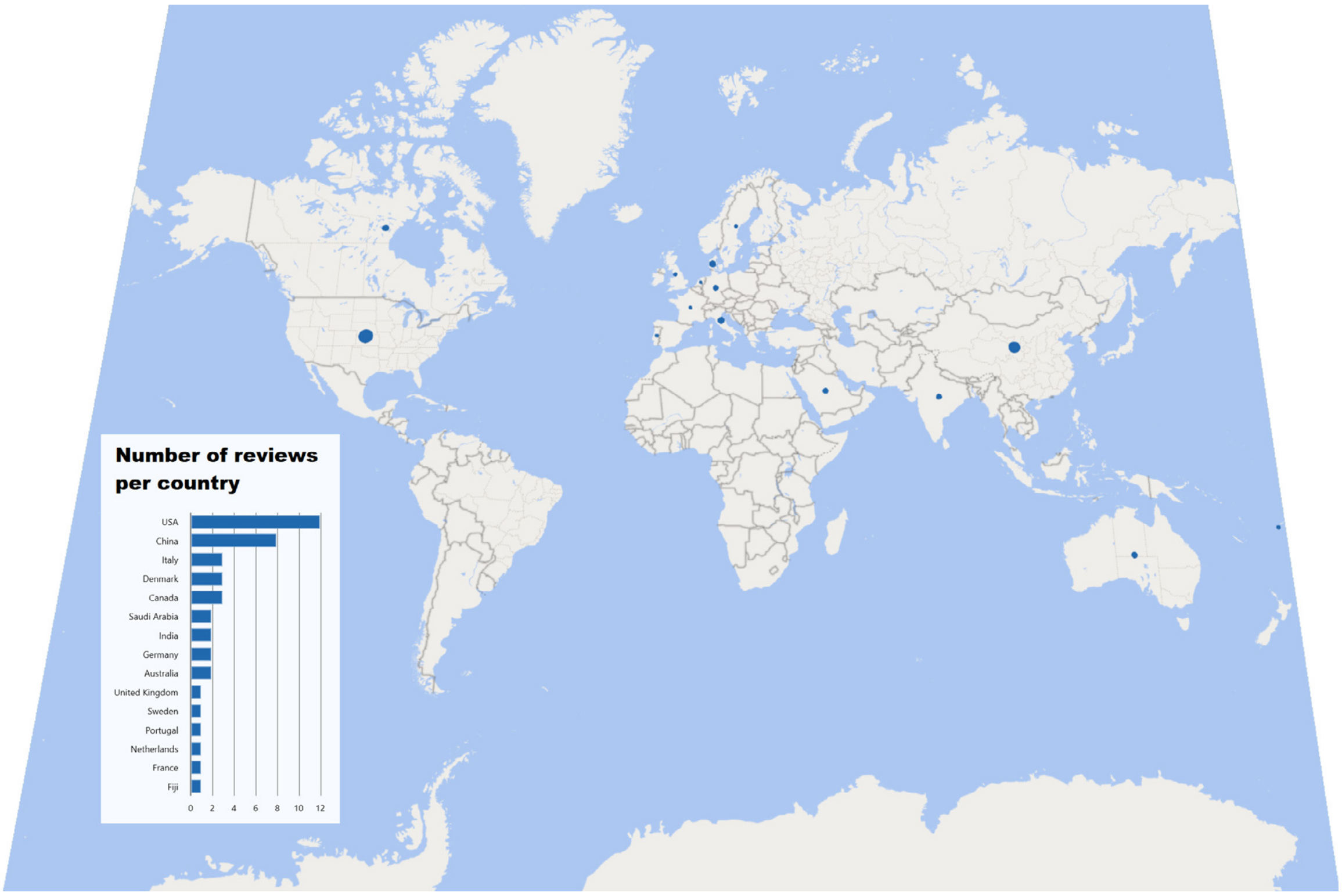

**Fig. 7.** World map showing the number of reviews per country according to the first author's affiliations.

deep learning toolkits, they can often be applied to new data without too much adaption.

- (5) The beforehand mentioned open access culture is promoted by publication venues, which require source code and data to be made openly available alongside the publication, like the Scientific Reports journal. This ensures reproducibility and verification by other researchers.
- (6) Furthermore, there are specific data journals, like Scientific Data or Data in Brief that provide venues to make medical datasets and data descriptors available to the research community. This makes it attractive to offer in-house datasets to the community (which is, first of all, a free service), because the data creators get an additional (citable) journal publication for their efforts.
- (7) Finally, deep learning is data-driven, which means it lives and dies by the amount of data it is fed, hence, the increasingly number of public medical databases, like the Cancer Imaging Archive or the Human Connectome Project, can be seen as very import driving forces behind the translation of deep learning into the medical domain.

It will be interesting to see what the future holds for us in the field of medical deep learning. Deep learning certainly already has an immense impact on the daily life of a large number of peo-ple via the countless applications that are based on this technique, such as, virtual personal assistants like Amazon's Alexa, Apple's Siri or Google's Now. However, as several real-life examples recently demonstrated, deep learning algorithms are not inerrant, as evidenced by tragic car accidents with self-driving cars, racist missclassifications of images, or the machine learning bot Tay from Microsoft that became (some kind of) a (virtual) sexist neoNazi [131]. Another relevant example is Google Photos, which identified two black persons as gorillas back in 2015 [132]. Someone could argue that a human raised and educated in a sexist or racist environment might also develop a similar behavioral attitude: Whether the algorithm did this willingly is more of a philosophical discussion. Interestingly, Google "*fixed*" the problem by removing and blocking the image categories "*gorilla*", "*chimp*", "*chimpanzee*" and "*monkey*". So, in summary, even leading technology companies face difficulties when it comes to ensuring that the output of data-driven algorithms do not lead to prejudices, racism or stereotypes of any kind. If we translate this issue to the medical area, where complex 3D volumes are used for (life-critical) clinical support, this is very significant. It should also be mentioned that tasks where deep learning outperformed humans have often been performed under *laboratory conditions*, with a fixed set of samples, not including real-life tests, or further weaknesses [133], and recent publications show how deep neural networks can easily be fooled [134].

In summary, we identified the following primary research gaps while analyzing the reviewed works:

- Almost none of the reported deep learning algorithms were incorporated into clinical workflows, mostly due to ethics and trust concerns ("How can we trust the neural network not to be wrong/biased, when we don't understand why it answers the way it does?"), making the testing and integration into clinical practice a prominent research gap.
- Along the same vein, research into more interpretable "Explainable AI" constitutes a large research gap that is particularly relevant to understand the underlying methods. And even more relevant to healthcare is an evidence-based medicine where an efficacy must be demonstrated empirically [135].
- A lack of well-annotated, multi-institutional, public datasets (particularly for medical disciplines using data other than radiographic images) was reported by most review authors, who also suggested that many individual papers reported the potential for increased performance based on more data. This research gap still exists today (early 2022), with particular relevance in niche disciplines or concerning rare diseases, where the data volume is low to begin with, but decreases in significance over time, as more and more such datasets and other techniques, like Federated Learning [136], become available.
- There exists a distinct lack of reliable standardized key performance indicators for deep learning methods in the field of medical research. Therefore, standardization of data, data acquisition and performance reporting represents an important facet of deep learning (albeit less of a research gap and more of a trend in the field).
- The tuning of model architecture, data processing and augmentations, and training hyperparameter choice appears to have a significant effect on the eventual performance of the model. However, due to the "black box" nature of most deep learning models, optimal choices in this regard are often difficult to ascertain. Optimization of this trial-and-error process represents a significant research gap, which is already an intensively discussed topic in the wider deep learning community.
- Only a few works cover multimodal data and the majority of works focus on single-modality data. However, physicians consider a multitude of resources when treating patients, which computer-assisted methods should also do and there should be a stronger focus on methods that can simultaneously process multimodal data [137].

## 5. Author's perspective

From a high-level point of view, and to formulate it provocatively, some tasks like medical image segmentation have already been solved over thirty years ago, as can be seen by the claims within the countless publications released in the past years. In addition, the entire computer vision field seems to move from a general hot topic to another one over the years, like deformable models in the late '80 s [138], graph-based approaches in early '00 s [139], and, finally, deep neural networks after 2010 [140]. This is also reflected by the sharp drop or rise of citations for these publications, depending on the addressed methodology. A more realistic picture of the feasibilities of the proposed works during these times may be biomedical challenges, where authors are encouraged to develop algorithms for a specific task [141], for example the very influential brain tumor segmentation (BraTS) challenge, about the automatic segmentation of brain tumors or our new AutoImplant challenge from 2020 [142], about automatic cranial implant design. The quantitative and qualitative evaluation results are often presented afterwards in a compact summary publication [143]. This definitely enables a more objective view on what is currently possible with the state-of-the-art methods (in this regard, also note the new BIAS guidelines for transparent reporting of biomedical image analysis challenges [144]), even though such challenges usually cannot replace a real evaluation in a clinical setting.

Finally, it should be mentioned that most medical deep learning applications are still in an early phase of development and have not yet found their way into real clinical practice. This stands in strong contrast to non-learning approaches, like those used in medical navigation systems for neurosurgery [145,146]. However, most computer science venues for dissemination, especially flagship venues, explicitly prefer and demand new algorithms, while works that focus on the applicability of existing methods to real, variable, and noisy clinical scenarios are nipped in the bud with the argument that they lack technical novelty. At the same time, to foster their status in academia, researchers commonly need to fulfill the expectations of selected publication venues. In many situations, world-leading experts and members of the MICCAI community have been expressing concerns about the practical usability of the research output, too often limited to ideal scenarios. It is not uncommon to hear criticism about that fact that even high-impact conference proceedings usually contain a huge number of tools and algorithms that are designed for ideal or limited scenarios and may be therefore inapplicable or sometimes unneeded. MICCAI fellow D. Shen (author of the very first review article in the field of medical deep learning according to our search strategy, see epub date in Table 2) summed up this issue in a recent public statement on LinkedIn [147]: "*In MICCAI field, people are studying same problems (sometimes ideal problems) with very similar methods for many years. Everyone claims their method is new (although mostly just simply borrowing from others). This is very serious issue, since people in this small academic field judge contributions of their works by themselves. If MICCAI people can just move a little bit out of their academic field, i.e., thinking more on real applications in clinical workflow, this issue can be largely avoided. We, as faculty, have more responsibility for changing this situation*". A step towards this direction could be that interdisciplinary and application-oriented venues encourage the involvement of a medical partner, including a statement of feasibility in the clinical practice. Furthermore, several interdisciplinary venues do not explicitly require any IRB approval statement, even if the manuscripts deal with clinical patient data (an exception here are publicly available datasets, but many works are still evaluated on private datasets provided by a medical partner, which also hinders an objective analyzing and reporting in reviews). In most medical venues, this is a standard requirement and submissions are rejected if the manuscript does not contain an official IRB or patient consent statement. Note that an approval from an ethics commission is also a pre-check for reasonability and should stop research endeavors that would harm the patient, for example by additional radiation exposure, or do not adhere to clinical workflows.

Nevertheless, and to pick up the "*not yet found their way into the real clinical practice*" and "*limited to ideal scenarios*" thoughts from before, some deep learning experts claim that just adding enough (training) data will automatically lead to perfect results. Contradictory to this opinion, cars have driven already millions of kilometers to acquire training data, but fully self-driving capabilities are still far away from being reliable, especially under different weather and light conditions. In this light, Tesla recently removed the "*full self-driving*" option from its car store on its website and Uber completely abandoned the development of self-driving cars. It should also be noted that in the medical field, such a massive amount of data will, in many cases, never be available. Certain pathologies are simply (and thankfully) not frequent enough, so even by collecting all the patient data for this pathology from the hospitals around the world and applying additional data augmentation methods [148], there still might not be enough for training powerful algorithms.

Nonetheless, there have been certain tasks where machine learning has undoubtedly outperformed humans already. Examples are Deep Blue [149] and AlphaGo [150] in games, where machine learning algorithms could even beat the best (known) human players around the world. However, these tasks have strong constraints by fixes rules on which algorithms can rely on. In contrast, medical tasks usually do not follow such rules and theoretically, unlimited possibilities exist. For example, a brain tumor [151] looks different for every patient in terms of shape, size, texture, etc. Another example is the human voice, with individual pitches and pronunciations, and further the inter-human variations when expressing different emotions [152]. In addition, algorithms can fall back on a massive database of pre-trained games and game moves, without any further uncertainty. Another example where deep learning works very well in practice is the automatic detection and analysis of car licenses. Despite several challenges and uncertainties, like different fonts, colors, languages, deformities, complex backgrounds, hazardous situations, speeding vehicles, occlusion, horizontal or vertical skew, blurriness, and illumination diversions [153], the recognition task still stays within a restricted rule set. Therefore, learning algorithms can be pre-trained, for example, by just going over the alphabet with variations, like changing the font, colors, adding some occlusion, etc. It should be kept in mind that still, vehicular license plate recognition is far from perfect.

In principle, deep learning is trying to mimic the human brain, especially the learning process of a human brain [154]. Equivalent to the fact that we cannot look into someone's brain with its thoughts or mindset, it is also not yet fully understood what is going on inside a deep neural network (even though we have access to all neurons and its connections, in contrast to a human's brain) [155]. Hence, it is as hard to predict exceptions and failures as seen in recent events, like car accidents, as it is to foresee human behavior and mistakes (even if there are, ironically, deep learning works that try to predict human behavior [156]). Trained neural networks with several layers and with a few hundred or a few thousand neurons are not understandable anymore in all detail [155]. This stands in strong contrast to pure engineering approaches, which can be understood in every detail. That makes the acceptance of such black box (some even call it Voodoo [157]) approaches, like deep learning, by the general population much harder. At this point, we want to refer the interested reader to the concept of disentanglement, which tries to make latent representations interpretable [158].

To conclude, deep learning is an exciting new field with a lot of potential, but not free of controversies. We believe that this first meta-review of medical deep learning reviews and surveys can provide a quick and comprehensive reference for scientists (or just interested readers) who want to get a high-level overview of this field, and maybe want to contribute and thus, accelerate the development in medical deep learning. Hence, the contribution of this systematic meta-review is sixfold:

- providing an overview of current deep learning reviews where a medical application plays the key role,
- arranging the researched works chronologically for a historical "*roten Faden*" (*red/common thread*) and picture over the years,
- extracting the overall number of referenced works and citations to give an impression of the research influence and footprints of the respective field,
- analyzing, exploring and highlighting the main reasons for the massive research efforts on this topic,
- conducting a comprehensive discussion of the current state-of-the-art methods in the deep learning area with achievements but also failures from other domains that should be avoided in the medical area,
- and providing a critical expert opinion and pointing out further controversies.

## Declaration of Competing Interest

The authors declare no competing financial interests.

## Acknowledgements

This work received funding from the Austrian Science Fund (FWF) KLI 678-B31: "*enFaced: Virtual and Augmented Reality Training and Navigation Module for 3D-Printed Facial Defect Reconstructions*", FWF KLI 1044: "*Instant AR Tool for Maxillofacial Surgery*" and the TU Graz Lead Project (*Mechanics, Modeling and Simulation of Aortic Dissection*). Moreover, this work was supported by CAMed (COMET K-Project 871132), which is funded by the Austrian Federal Ministry of Transport, Innovation and Technology (BMVIT), and the Austrian Federal Ministry for Digital and Economic Affairs (BMDW), and the Styrian Business Promotion Agency (SFG). Furthermore, we acknowledge the REACT-EU project KITE (Plattform für KI-Translation Essen). Finally, we want to make the interested reader aware of our medical image processing framework *Studier-Fenster* (www.studierfenster.at) [159], where medical deep learning approaches can be tried out in a standard web browser.

## Supplementary materials

Supplementary material associated with this article can be found, in the online version, at doi:10.1016/j.cmpb.2022.106874.